\documentclass{aastex}

\usepackage{psfig,emulateapj5,times,mathptm}

\usepackage{endfloat}

\received{}
\accepted{}

\begin{document}

\title {ESI, a new Keck Observatory echellette spectrograph and imager}


\author{A.I. Sheinis,  M. Bolte, H.W. Epps, R.I. Kibrick, J.S. Miller, M.V. Radovan \altaffilmark{1},\\ B.C. Bigelow,  B.M. Sutin\altaffilmark{2}}



\altaffiltext{1}{UCO/Lick Observatory, University of California Santa Cruz, Interdisciplinary Science Building, Santa Cruz, CA 95064\\
\email{sheinis@ucolick.org, bolte@ucolick.org, epps@ucolick.org, kibrick@ucolick.org, miller@ucolick.org, mvr@ucolick.org}}
\altaffiltext{2} {Carnegie Observatories, 813 Santa Barbara Street, Pasadena, California 91101\\
\email{bigelow@ociw.edu, sutin@ociw.edu}}

\begin{abstract} The Echellette Spectrograph and Imager (ESI) is a
multipurpose instrument which has been delivered by the Instrument
Development Laboratory of Lick Observatory for use at the Cassegrain
focus of the Keck II telescope. ESI saw first light on August 29,
1999. ESI is a multi-mode instrument that enables the observer to
seamlessly switch between three modes during an observation. The three
modes of ESI are:  An R=13,000-echellette mode;  Low-dispersion
prismatic mode;  Direct imaging mode.  ESI contains a unique
flexure compensation system which reduces the small instrument flexure
to negligible proportions. Long-exposure images on the sky show FWHM
spot diameters of 34 microns (0\farcs34) averaged over the entire
field of view. These are the best non-AO images taken in the visible
at Keck Observatory to date. Maximum efficiencies are measured to be
28{\%} for the echellette mode and greater than  41{\%} for
low-dispersion prismatic mode including atmospheric, telescope and
detector losses.  In this paper we describe the instrument
and its development. We also discuss the performance-testing and some
observational results.

\end{abstract}

\keywords{ ESI, instrumentation, spectrograph, imager, echellette, Keck Observatory, spectroscopy, astronomy\\[5cm]}



\section{Overview}

The Echellette Spectrograph and Imager (ESI) is a recently
commissioned instrument available for use at the Keck II telescope of
the W.M. Keck Observatory.  ESI was built by the Instrument
Development Laboratory of UCO/Lick Observatory at the University of
California, Santa Cruz campus. The ESI instrument has three modes.
The 160-mm diameter collimated beam can be sent directly into the
camera for direct imaging, through a prism disperser for low
resolution spectroscopy, or to an echellette grating with prism
cross-dispersion for moderate-resolution, large-wavelength-coverage
spectroscopy.

An all-refracting ten-element all-spherical camera and a single ($2K
\times 4$K by 15-micron) CCD detector are used without modification
for all three modes.  The direct-imaging mode has a ($2\farcm0 \times
8\farcm0$) field of view (f.o.v.) with 0\farcs15 pixels.  The
low-dispersion prism-only mode (LDP) has a reciprocal dispersion of 50
to 300 km/sec/pixel, depending on wavelength.  This mode can be used
with an 8\farcm0-long slit or in a multi-slit mode with user-made
slitmasks.  The higher-dispersion echellette mode gives the entire
spectrum from 0.39 to 1.09 microns with a 20\farcs0 slit length, in a
single exposure. The reciprocal dispersion in echellette mode varies
from 9.6 km/sec/pixel to 12.8 km/sec/pixel.  Three five-bay filter
wheels provide locations for up to eight different full-field filters,
three multi-object slit masks and the standard complement of five
echellette slits in a single bay. Each wheel also contains a single
open bay.

The instrument was delivered to Keck II in July of 1999 and saw first
light on August 29, 1999. In its first two years of operation it has
been scheduled for an average of 111 nights per year with instrument-reliability losses accounting for less than 2\% of this time. In the imaging mode,
ESI routinely delivers remarkable image quality. Long exposures with
FWHM values $<0\farcs45$ have been commonly obtained.

The Principal Investigator (PI) of the project was Joe Miller, Director
of Lick Observatory.  ESI also had three Co-PI's; Michael Bolte, Puragra
Guhathakurta, and Dennis Zaritsky. The optical designers were Brian
Sutin and Harland Epps.  David Cowley was the project manager.
Initially (11/95 - 7/97), Bruce Bigelow was the  project scientist.
After Bigelow left for Carnegie Observatory and for the remainder of
the project, the  project scientist responsibilities were split
between  project engineers Matt Radovan and Andrew Sheinis.

The principal niche envisioned for ESI was very high-throughput
moderate-resolution optical spectroscopy with wide wavelength coverage
in a single exposure. Examples of the science programs which drove the
choice of resolution are kinematic and abundance studies of
distance galaxies, detailed abundance determinations for stars as
faint as V$=22$ (which includes the bright giants in Local Group
galaxies) and absorption-line studies along the lines-of-sight to
QSOs. The low-dispersion/multi-object and direct-imaging capabilities
were added after the echellette design was conceptualized. In the
low-dispersion mode, ESI is likely the most powerful spectrograph in
operation for obtaining low-resolution spectra of extremely faint
objects. Redshift determinations and spectral energy distribution measurements for
very faint galaxies or low-luminosity stars in the Galaxy are 
possible to V$ \approx 25$. 
Examples of first-light science can be seen in  Becker et al. (2000) 
and \cite{PGW01}.

This paper serves as a comprehensive overview of the instrument.  For
details about the individual subsystems, the reader is directed to the
other articles in the literature describing various aspects of ESI:
\cite{KMNRSS00}; \cite{SNJR99}; \cite{SSESHL99}; \cite{RBNS98};
\cite{SNR98}; \cite{EPP98}; \cite{EM98}; \cite{SUT97}; \cite{BN98};
\cite{SMBS00}. For observers planning to use ESI, the Keck Observatory
WWW site for the instrument \url{http://www2.keck.hawaii.edu:3636/realpublic/inst/esi/esi.html} contains detailed specifications and descriptions of
operational procedures.

The wavelength range for ESI was chosen to lie between 0.39 microns
where the best glasses for the visual range become acceptably transparent, up
to 1.1 microns where the efficiency of silicon CCD detectors approaches zero. The f.o.v. imaged by ESI is a rectangle located
5\farcm0 off-axis from the telescope axis.  It is 2\farcm0
wide in the radial direction by 8\farcm0 long. With this f.o.v., an
on-axis (nearly) parabolic collimator, imaging an off-axis field was
required to minimize off-axis aberrations.  The 2286.0-mm focal length of the
collimator was chosen to be the longest that would fit into the
Cassegrain module envelope, giving a 160.0-mm diameter beam. After the collimator, the optical elements used in the instrument are:  Two flat fold mirrors;  Two prisms;  A
 grating;  The camera and shutter assembly;  The Dewar.  

The requirements for the spectrograph optical layout are:

\begin{itemize}

\item The echellette spectral format must fill the CCD chip efficiently;

\item The orders must not overlap when a  20\farcs0 slit is used;

\item The individual slit images must be quasi-parallel over the entire detector;

\item The incoming beams should have minimum vignetting at the camera mouth.

\end{itemize}

\textbf{Mode 1: Echellette in near-Littrow configuration}\\
\\
The primary mode of the spectrograph uses a catalogue echellette
grating (Spectronics $204 \times 204$ mm, 175 lines/mm, 32.3-degree
blaze angle), to achieve a spectral resolution of 13,000 at the
minimum two pixel slit-width = $0\farcs3 $. This mode disperses the
light from the $20\farcs0$ slit into 10 orders, from order-6 in the
red to order-15 in the blue.  The minimum order separation is
$22\farcs0$.  Table~\ref{table:orders} shows diffraction orders,
wavelength ranges, reciprocal dispersions and rms image diameters for
the echellette mode. The rms image diameters were calculated as a
function of wavelength for all of the echellette orders.  The
calculation includes aberrations from the entire optical train.  The
diameters were calculated by apodizing the irregularly-shaped Keck II
primary mirror and by assuming perfect seeing. 

\begin{table}
\begin{center}
\caption{Predicted Performance, Echellette Mode}
\vskip 8pt
\begin{tabular}{llcclc}\hline 
\hline
\multicolumn{1}{c}{Order} & range$(\mu)$ & D$_{rms}(\mu)$ & D$_{rms}$(arcseconds)
 &  \AA/pix & Arcseconds/pix \\
\hline
15 & 0.393 $\cdots$  0.419 & 12.01 &  0.122 & 0.15 & 0.127  \\
14 & 0.420 $\cdots$ 0.451 & 16.03 &  0.163 & 0.17 & 0.133 \\ 
13 & 0.451 $\cdots$ 0.486 & 19.58 & 0.199 & 0.18 & 0.141  \\
12 & 0.487 $\cdots$ 0.529 & 19.15 &  0.195 & 0.19 & 0.148 \\ 
11 & 0.529 $\cdots$ 0.581 & 17.20 &  0.175 & 0.21 & 0.153  \\
10 & 0.581 $\cdots$ 0.640 & 17.86 &  0.182 & 0.23 & 0.159  \\ 
 9 & 0.640 $\cdots$ 0.715 & 22.24 &  0.226 & 0.26 & 0.164  \\
 8 & 0.715 $\cdots$ 0.813 & 25.46 &  0.259 & 0.29 & 0.171  \\
 7 & 0.813 $\cdots$ 0.937 & 24.19 &  0.246 & 0.33 & 0.177  \\
 6 & 0.937 $\cdots$ 1.093 & 15.93 &  0.162 & 0.39 & 0.183  \\
\hline 
\label{table:orders}
\end{tabular}
\end{center}
\end{table}

\begin{figure}
\plotone{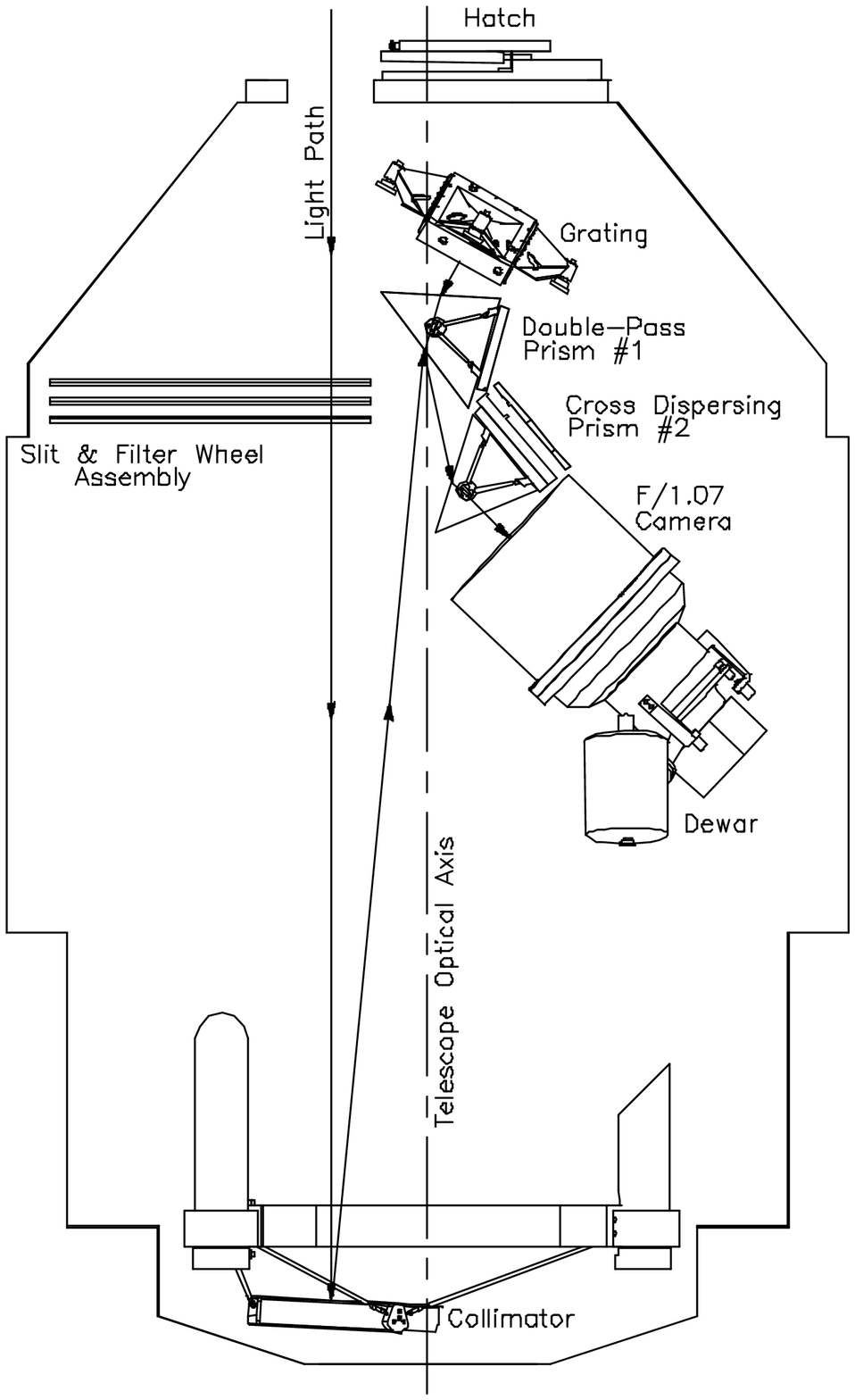}
\caption{ Schematic layout of Echellette mode. This figure shows the physical layout of the
spectrograph in echellette mode.  The collimated light from the slit
is pre-cross-dispersed by the first prism before it reaches the
echellette grating, which is used nearly in Littrow
for high efficiency.  The grating-dispersed light is once again cross-dispersed
by a second pass through the first prism and then passes
through the second prism before being imaged onto the detector by the camera. }
\label{fig:echlayout}
\end{figure}

Figure ~\ref{fig:echlayout} shows the physical layout of the
spectrograph in echellette mode.  The collimated light from the slit
is pre-cross-dispersed by the first prism before it reaches the
echellette grating, which is used nearly in Littrow
for high efficiency.  Although the spectrograph was designed
specifically for this grating, other gratings could be manually
installed.  The grating-dispersed light is once again cross-dispersed
by a second pass through the aforementioned prism. It then passes
through the second prism before being imaged onto the detector by the
underfilled f/1.07 camera.

\begin{figure}
\plotone{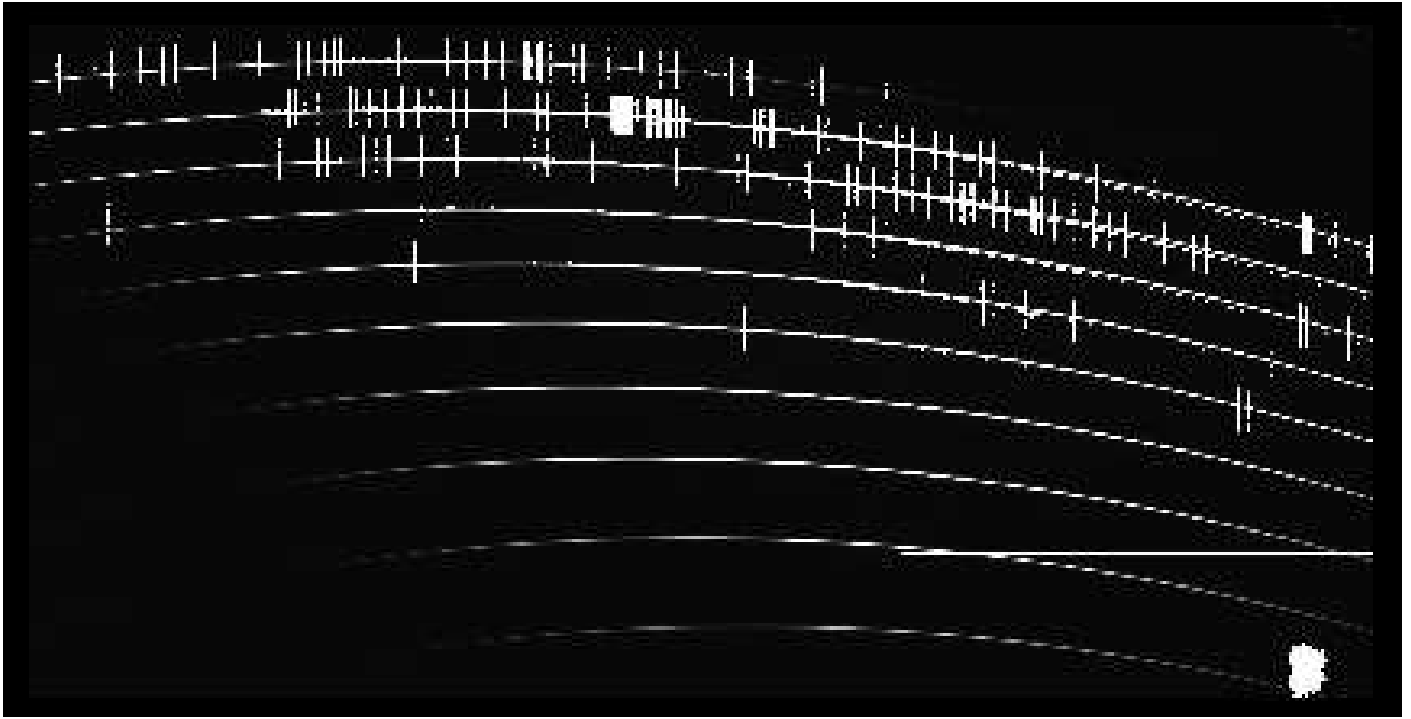}
\caption{ Echellette spectrum. This figure shows an unprocessed echellette spectrum of a
reference star. Note the strong night-sky emission lines. The entire spectral range from 0.39 to 1.09 microns is
imaged without gaps.  Orders run from 6 at the top to 15 at the bottom
with red to the right. Note the two significant cosmetic defects in
the lower-right quadrant.  A glowing pixel at row 85 of column 3895
contaminates a (10 by 10) pixel area, while 18 adjacent bad rows on
the rightmost $\onethird$ of the device affect order 14 from 4460 to
4565 \AA.  However those wavelengths are available outside of the free
spectral range in order 13. (Note: this image has been rotated 90 degrees in software, hence the CCD serial register runs vertically along the left edge of this figure.)}
\label{fig:echspec}
\end{figure}

Figure ~\ref{fig:echspec} shows an unprocessed echellette spectrum of a
reference star. The entire spectral range from 0.39 to 1.09 microns is
imaged without gaps.  Orders run from 6 at the top to 15 at the bottom
with red to the right. Note the two significant cosmetic defects in
the lower-right quadrant.  A glowing pixel at row 85 of column 3895
contaminates a (10 by 10) pixel area, while 18 adjacent bad rows on
the rightmost $\onethird$ of the device affect order 14 from 4460 to
4565 \AA.  However those wavelengths are available outside of the free
spectral range in order 13.

The echellette orders are curved primarily by the anamorphic
distortion in the prisms. They show a reasonably constant wavelength
solution as a function of slit height. These wavelength solutions were
used to calculate the residual tilts of the slit images. These tilts range from $0.8 \pm 0.2$ in order 6 to $-1.3 \pm 0.7$ in order 15. Detailed
plots of slit tilt with wavelength are available in \cite{SUT97}.

\begin{figure}
\plotone{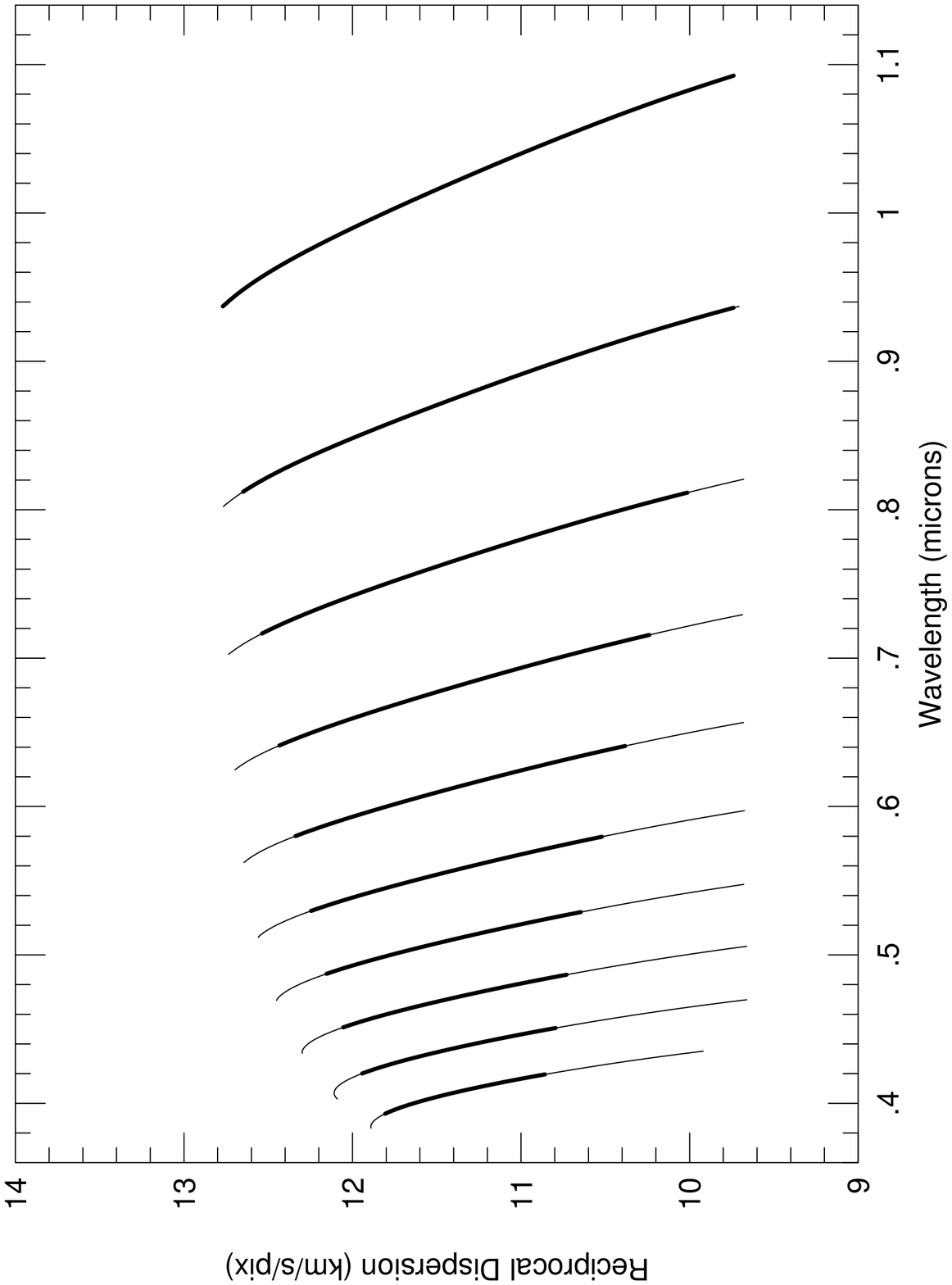}
\caption{ The reciprocal
dispersion as a function of wavelength for each order in the
echellette mode The darkened portions of the curves show the free-spectral range, while the lighter portions show the full extent of each order. }
\label{fig:echdispersion}
\end{figure}

 Figure ~\ref{fig:echdispersion} shows the reciprocal
dispersion as a function of wavelength for each order in the
echellette mode.  The average reciprocal dispersion is about 11.4
km/sec/pixel.

\textbf{Mode 2: High-efficiency, low-dispersion configuration}\\
\\
The high-efficiency, low-dispersion mode is achieved by modifying the
optical path in the echellette mode (Figure~\ref{fig:echlayout}). This
is done by inserting a flat-silvered mirror in front of the echellette
grating such that the beam will bypass the grating and be dispersed
exclusively by the two prisms. The slit direction is then rotated by
90 degrees to accommodate the new dispersion direction.  The main
strength of this mode is the high throughput, as the silvered mirror
has higher efficiency than the aluminized grating.

Since the exit pupil of the collimator is near the prisms, f.o.v. in this mode is not significantly restricted by vignetting at
the prisms.  Thus an $8\farcm0$ long slit may be used.  This maximum slit
length is where the images begin to get ``soft.''  With a field this
long, up to 50 slitlets could be cut into a slit mask within the
$2\farcm0 \times 8\farcm0$ viewing area.  

\begin{figure}
\plotone{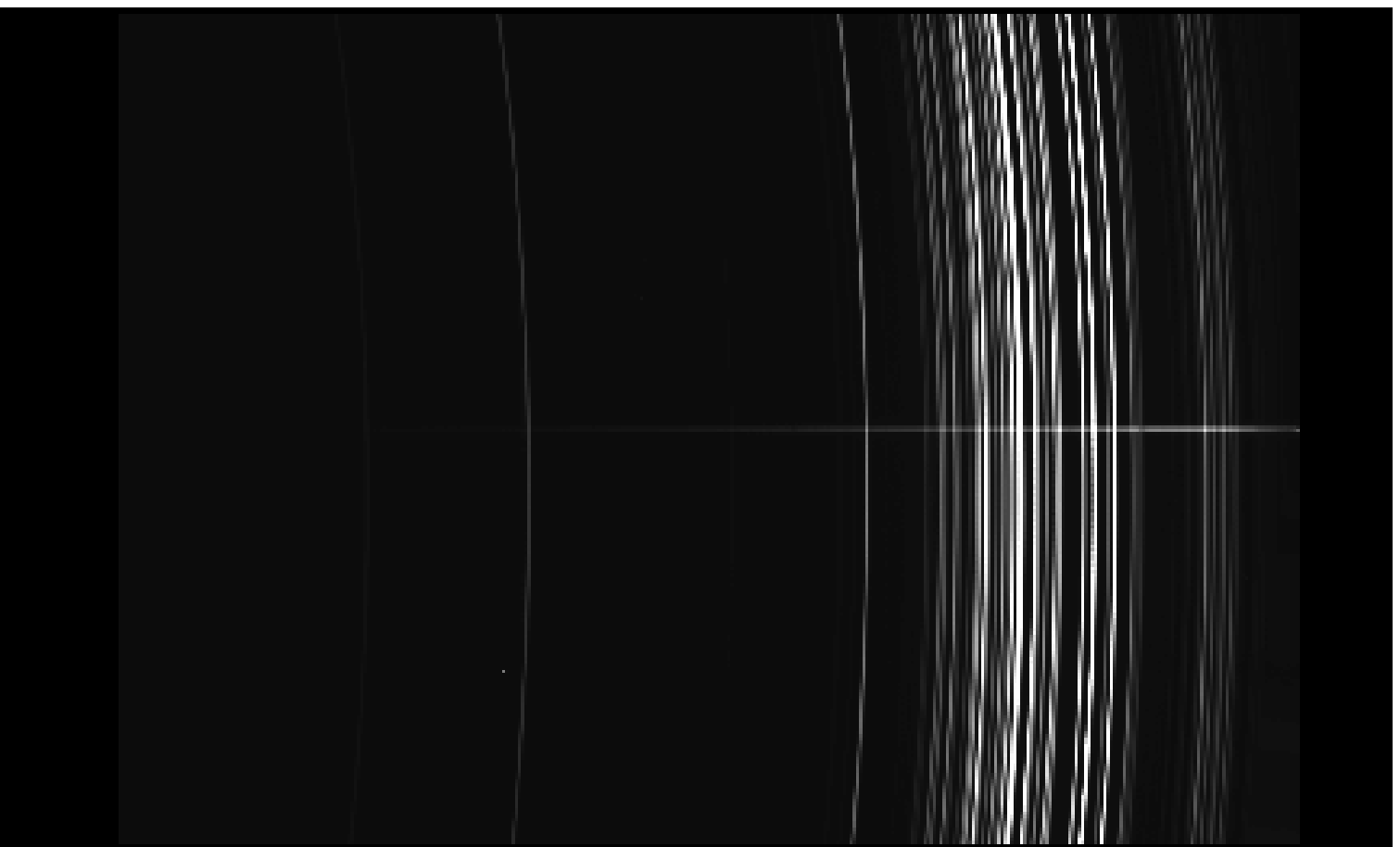}
\caption{ The spectrum of a reference star taken
in the high-efficiency, low-dispersion mode  with ESI. This image has been rotated with software so as to align the apparent dispersion direction approximately along the detector row direction.  Note the curved sky lines due to the prismatic anamorphic distortion. }
\label{fig:lowdspec}
\end{figure}

 Figure~\ref{fig:lowdspec} shows the spectrum of a reference star taken
in this mode with ESI. This image has been processed using the Information data language (IDL) so as to align the apparent dispersion direction approximately with a single row.  Note the curved sky lines due to the prismatic anamorphic distortion.

\textbf{Mode 3: Direct imaging configuration}\\
\\
The direct imaging mode is achieved by modifying the
optical path in the echellette mode (Figure~\ref{fig:echlayout}). This
is done by translating the large post-dispersing prism out of the beam and replacing it with a flat silvered mirror such that the beam will bypass all the dispersing elements and be sent directly to the camera, producing a $2\farcm0 \times 8\farcm0$ f.o.v..  The slit-wheel is generally turned to the open position, while the standard filter options, Johnson (B), Johnson (V), Spinrad (R), Gunn (I) are available only in a $2\farcm0 \times 2\farcm0$ format. The third filter wheel is available for user-supplied filters up to the $2\farcm0 \times 8\farcm0$ format.

\section{ Opto-Mechanical Subsystems}

The design of ESI was divided into several natural opto-mechanical subsystems, namely:
 The collimator; The prisms;  The optical flats including the
grating;  The camera assembly;  The detector and controller;  The slit
and filter wheel assembly;  The calibration system;  The guider
assembly. We now briefly discuss each of those subsystems. References are listed for readers interested in more detail.

\subsection{Collimator}
 
\begin{figure}
\plotone{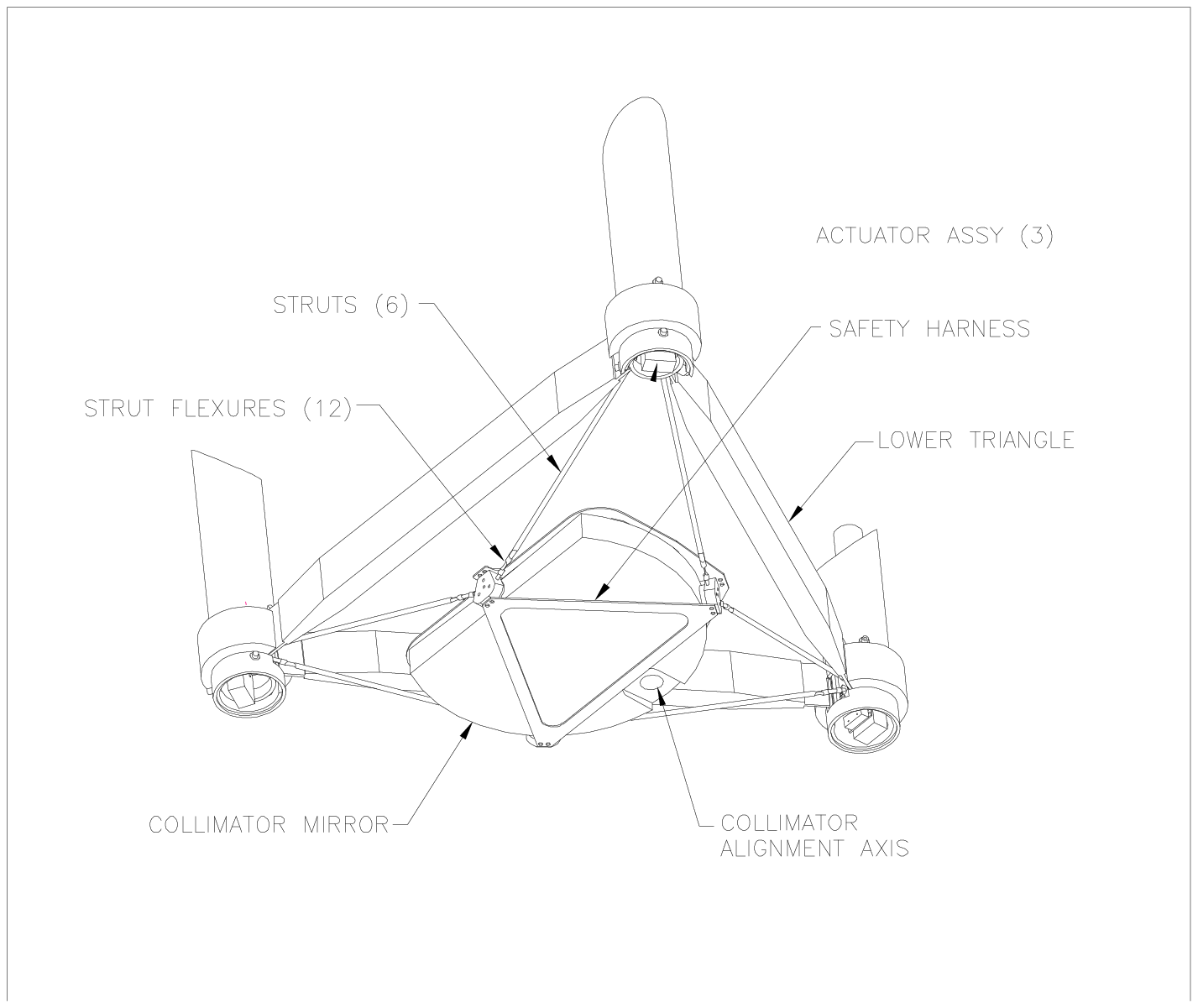}
\caption{ ESI Collimator system design, which is actively
articulated in piston, tip and tilt to provide the focus and the
flexure control for the instrument.  The collimator mirror is 580 mm
in diameter with a focal length of 2286 mm. }

\label{fig:collimator}
\end{figure}

ESI uses a novel collimator system design, which is pictured in
Figure~\ref{fig:collimator}, and is actively articulated in piston,
tip and tilt to provide the focus and the flexure control for the
instrument. The details of this system are described in \cite{RBNS98}.
The collimator mirror is 580 mm in diameter to accommodate the
$2\farcm0 \times 8\farcm0$ f.o.v.. It has a focal length of 2286
mm. The range of focus was selected to be $\pm 25$ mm which gives
out-of-focus star-image diameters of 2\farcs4.  The collimator mirror
is optimally constrained by a space-frame structure.  The flexure
compensation is accomplished by applying calibrated active correction for system gravity flexure in an open-loop fashion, to the collimator.
 
 
The flexural requirement for any two-hour observation was that the spectra
 should remain stable on the detector to within $\pm 0\farcs04$
 (peak-to-valley) with (open-loop) flexure compensation. At the final
 imaging scale of $ 0\farcs153$ per pixel, this requirement
 corresponds to $ \pm \frac{1}{4}$ pixel.  In order to achieve the
 required flexure compensation precision, collimator tilt increments
 must be controlled in 0\farcs01 units upon the sky, corresponding to
 collimator tilts of 0\farcs327.  With actuator separations of 1200
 mm, that corresponds to actuator motions of 1.65 $\mu$. In order to
 achieve the desired correction precision, hysteresis and slip levels
 were maintained below this value. 
 

\subsection{ Cross-Dispersion Prisms} 


ESI uses two large (25 kg), $50.0^\circ$-angle prisms for
cross-dispersion. The optical requirements for the prisms were high
transmission, small index inhomogeneity, a large ratio of red
dispersion relative to blue dispersion while maintaining as high an
index as possible. Fused silica was considered, however its relatively
low index would have required $55.0^\circ$-prisms which would have
complicated the optical and coating designs.  Ohara BSL7Y was selected
for the prisms as its optical properties provide an optimal
combination of the requirements listed above.

Various index-inhomogeneity models for the prisms were computed and
ray traced. The results of this modeling suggested that the index
homogeneity must be constant to approximately $\pm 10^{-5}$. Our
contract with the vendor did not guarantee this level of
precision. However, we were able to request fabrication on a ``best
effort'' basis. Interferograms of the finished prisms taken in
transmission, along with the final image quality tests, show that the
fabricated prisms are homogeneous to approximately $\pm 2$ x $ 10 ^{-6}$.

 The prisms are mounted within the spectrograph using
determinate structures. These structures  are extremely rigid for their weight because  mount material is
used in compression or tension only. The mounts are
designed such that no moments can be imparted at the mount-to-glass
interfaces, thereby insuring that the glass is isolated from any
mechanical stresses developed inside the instrument structure. The prisms and their associated mounting mechanisms are described in \cite{SNR98}.

\subsection{Kinematic mirror mechanisms}

ESI uses two moderate weight (10kg) translating fold mirrors to switch
between modes. The scientific requirements for the spectrograph are
such that the spectra may move by no more than $ \pm 0\farcs04$ on the
sky ($\frac{1}{4}$ pixel) in switching between modes and back again, so
as to accommodate the use of a single set of standard reference
spectra and flat-field data between different exposures. This places a
$\pm 1\farcs3$ allowable tip and tilt repeatability on the mirrors.  A
novel locating mechanism design was developed to achieve the
repeatability needed for these mirrors. This mechanism is described in
detail in \cite{SNJR99}.

\subsection{Camera Subsystem}

\begin{figure}
\plotone{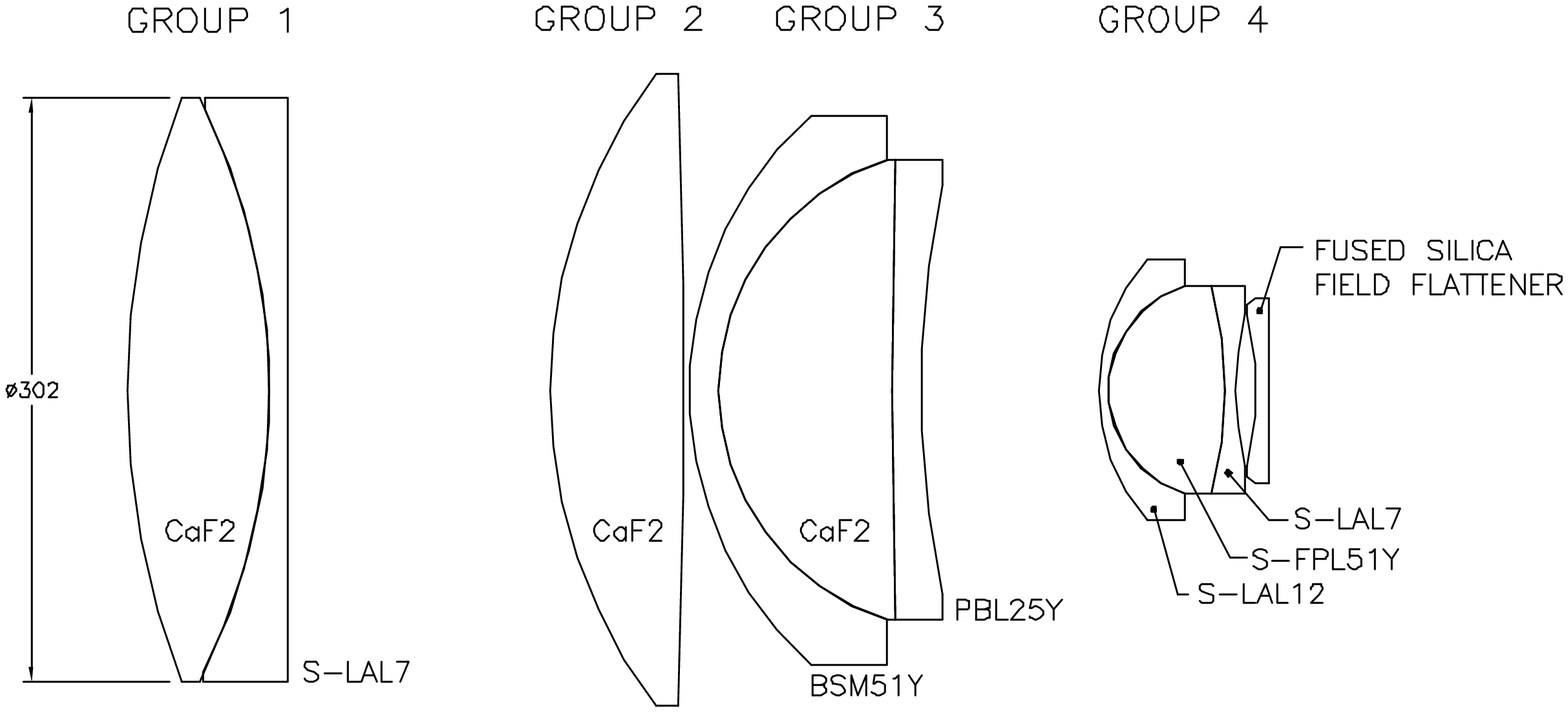}
\caption{ The ESI camera design is all-spherical and it includes three
large CaF$_2$ lenses. The elements in groups \#1 and \#3 are
optically coupled with a fluid (Cargille laser liquid Type 5610
{$n_{D}=1.5000$}) to minimize internal reflections. The elements in
group \#4 are greased together with Dow Corning Q2-3067 optical
couplant. }

\label{fig:camopticaldesign}
\end{figure}

The camera optical design is shown in Figure~\ref{fig:camopticaldesign}. 
It consists of ten lens elements in five lens
groups.  The camera has an effective focal length of 308 mm. It has an
entrance aperture diameter of 287 mm and a final plate scale of 97.7
microns/arcsec on the sky. The collimated beam diameter is
approximately 160 mm. The camera's effective f/ratio in imaging
mode is thus f/1.93 and slightly faster in spectroscopic modes due to
anamorphism.

The camera design is all-spherical and it includes three large CaF$_2$
lenses. Group \#1 is a doublet, group \#2 is a CaF$_2$ singlet, groups \#3
and \#4 are triplets while group \#5 is the field flattener/Dewar
window. The elements in groups \#1 and \#3 are optically coupled with a
fluid (Cargille laser liquid Type 5610 {$n_{D}=1.5000$}) to minimize
internal reflections. The elements in group \#4 are greased together
with Dow Corning Q2-3067 optical couplant. The six larger elements
were fabricated by TORC (Tucson, AZ) and the 4 smaller elements were fabricated by
Cosmo Optics Inc., (Middletown, NY). Broad-passband AR coatings were
applied by Coherent Auburn Division (Auburn, CA).

The ESI optical design is complicated by the fact that a wide variety of
pupil anamorphic factors and effective entrance pupil distances is
presented to the camera{'}s entrance aperture by the three
operating modes. (In practice, the camera design was slightly compromised in
the imaging and LDP modes so as to favor the echellette mode). Nevertheless, the
echellette mode remains the most severe test of system image quality.

In order to produce detailed sensitivity analysis, individual lens
elements were perturbed axially, radially, and in tip/tilt within the
computer model.  The perturbed camera model was ray traced within the
ESI spectrograph numerical model in echellette mode at 50 different
wavelengths, covering all 10 orders.  The perturbed system was
compared to the unperturbed system, in order to determine the
sensitivity coefficients of the camera. The worst centroid location
change and worst increase in the rms spot-size diameter (D$_{rms}$)
were noted.  A similar analysis was performed for the lens
groups. After moving entire groups, we calculated the resulting
centroid motion and change in D$_{rms}$.

These sensitivities were used to estimate manufacturing and assembly
tolerances for the mechanical components such that the expected
summation of image errors (added in quadrature) was comparable to the
residual aberrations in the optical design. In addition, the
sensitivity data allowed setting of a lens-motion tolerance in
terms of the total allowable image motion for the different gravity
orientations.

\subsubsection{Camera Opto-Mechanical Design}

\begin{figure}
\plotone{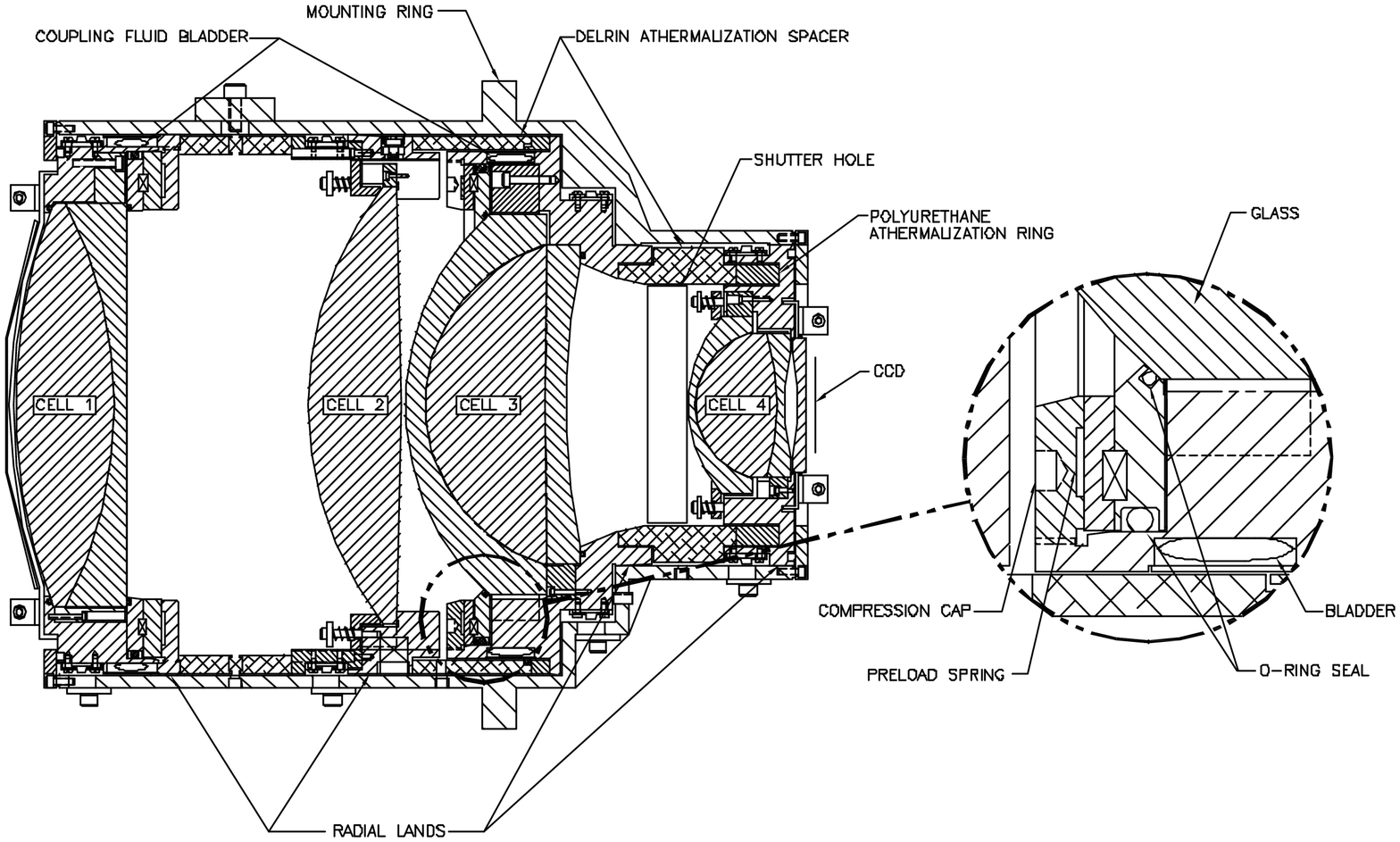}
\caption{Camera mechanical system (CMS) consists of four cells
 supported within a single large barrel.  The cells locate
 radially against locating surfaces (lands) on the barrel inner wall
 and axially against athermalizing spacers. Each cell consists of an
 aluminum housing, radial athermalizing spacers and a compressive
 preload spring. Each of the two oil-coupled cells also contains an
 oil-sealing and oil-reservoir system. }

\label{fig:cammechanicaldesign}
\end{figure}

 The camera mechanical system (CMS) for ESI was designed by J. Alan
 Schier Co., ( La Crescenta, CA) and fabricated by Danco
 Machine, DPMS Inc., (Santa Clara, CA). The CMS is shown in Figure~\ref{fig:cammechanicaldesign} . It
 consists of four cells supported within a single large barrel. The
 mass of the entire camera assembly, consisting of the CMS and the
 lens elements, is approximately 125 kg. The individual cells with
 their respective lens elements have masses ranging from 5 to 30
 kg. The cells locate radially against locating surfaces (lands) on
 the barrel inner wall and axially against athermalizing spacers. Each
 cell consists of an aluminum housing, radial athermalizing spacers
 and a compressive preload spring. Each of the two oil-coupled cells
 also contains an oil-sealing and  oil-reservoir system. The lens element
 spacings within the oil-coupled cells are maintained by 0.10-mm Mylar
 spacers, placed between the lens elements at their edges.

To accommodate the optical coupling fluid, cells \#1 and \#3 contain an
o-ring seal at each glass-aluminum interface. The o-rings are
compressed to the minimum nominal-dynamic-seal specification (8\%).
The o-rings are compressed such that the optic locates directly
against the aluminum land and the o-ring carries most of the load,
while providing a seal.

         A polyurethane bladder is used to accommodate volumetric
changes of the coupling fluid within each cell. Possible reactivity of
the optical coupling fluid with polyurethane and other substances was
tested by \cite{HLF99}. They found no reactivity with the polyurethane
bladder material.

The major features of the opto-mechanical design and the assembly of the camera are detailed in \cite{SSESHL99}.

\subsubsection{Camera Bench Testing and Performance Analysis}

After assembly of the camera, numerous optical performance tests,
including interferometric testing, polychromatic-point-spread function
measurements and back-focal-distance measurements were carried out.
The results of these tests compared well with the predictions from the
lens design. For example, the polychromatic (Wratten 54) rms spot
diameter on-axis was measured to be 13.7 microns, as compared to a
predicted rms spot diameter of 12.5 microns.  These tests are
discussed in detail in \cite{SSESHL99}.


\subsection{Detector and Controller} 
 
The ESI detector (device W62C2) is a MIT/Lincoln Labs CCID20
high-resistivity CCD . For details see \cite{BGLCYCLLT00}.  It uses readout amplifiers at both ends of the serial register.  Measured deviations from flatness are 18
microns peak to valley at the normal operating temperature of $-120^\circ.0$
 C.  At that temperature, the on-chip amplifier read noise as
measured in the laboratory is 2.3 electrons for one amplifier and 2.6
electrons for the other.  When the CCD Dewar is mounted in ESI, the
ESI instrument is mounted on the Keck II telescope, all of the
motorized stages within ESI are powered up and operating and the telescope is
tracking,  the measured readout noise for each amplifier increases by
0.3 electrons.
 
Operating at $-120^\circ.0$ C, the dark current is 2.1
electrons/pixel/hour in non-MPP mode.  At that temperature, the
various wavelength-dependent specifications for the CCD are shown in
Table ~\ref{table:fringing}. Note, these MIT/LL devices are not capable of
MPP mode operation.

\begin{table}
\begin{center}
\caption{CCD QE and Fringing}
\vskip 8pt
\begin{tabular}{ccc}
\hline 
\hline
\multicolumn{1}{c}{Wavelength} &  QE (\%)  &  Fringing (\%) \\
\hline 

3200 \AA      &      10.2 &	\\
4000 \AA      &      60.8 &	\\
5000 \AA      &	     82.4 &   \\
6000 \AA      &      80.3 &	\\
7000 \AA      &      77.1 &	\\
8000 \AA      &      68.9 &      12.0 \\
9000 \AA      &      45.0 &      15.0 \\
10000 \AA     &      11.3 &      30.0 \\

\hline 
\label{table:fringing}
\end{tabular}
\end{center}
\end{table}

The CCD controller for ESI is a San Diego Sate University
second-generation device (SDSU2) developed by \cite{LBE98}.
It provides both a low-gain and a high-gain mode of operation.  While
the pixel full-well of this CCD is 105,000 electrons, the achievable  
full-well is limited by the 16-bit resolution of the analog to digital
conversion.  Table ~\ref{table:full-well} shows the achievable full-well as a function of the selected gain.

\begin{table}
\begin{center}
\caption{Achievable Full-Well Versus Gain}
\vskip 8pt
\begin{tabular}{cccc}
\hline 
\hline
\multicolumn{1}{c}{Mode } &  Gain   &  Achievable & \%  of Pixel Full-Well \\
\hline

low gain    &    1.30 e-/DN    &   84,500 e-    &    80 \\
high gain    &    0.52 e-/DN    &   31,000 e-    &    30 \\
 
\hline 
\label{table:full-well}
\end{tabular}
\end{center}
\end{table}

The controller also provides a software-selectable choice of three  
different readout speeds.  However nearly all observations are performed
using the fastest of the three speeds.  Table ~\ref{table:pixel timing}  provides the details of
the per-pixel timing for the fastest and slowest speeds while Table ~\ref{table:CTE} shows
the serial charge transfer efficiency for this CCD as a function of the 
selected readout speed.  Note that the parallel charge transfer efficiency for this device is
0.9999999 (virtually perfect) at $-120^\circ.0$ C for all readout speeds and the
readout noise is the same at all readout speeds.  While the slow speed readout
mode is seldom used, it does provide for nearly perfect serial charge transfer efficiency at 
the cost of significantly increased readout time.

\begin{table}
\begin{center}
\caption{Per Pixel Timing Versus Readout Speed}
\vskip 8pt
\begin{tabular}{lccccc}
\hline 
\hline
\multicolumn{1}{c}{Readout Mode } & Sampling Time    & Serial  & Pixel & Pixel & Pixel Frequency\\
\multicolumn{1}{c}{ } & (Baseline+Video)    & Width &Overlap  &Period & Per Amplifier \\
\hline 
fast      &    1 $\mu$ sec + 1 $\mu$ sec  &  3 $\mu$ sec  & 1 $\mu$ sec  &  7.2 $\mu$ sec  & 139 Kpixels/sec \\
slow       &    5 $\mu$ sec + 5 $\mu$ sec  & 17 $\mu$ sec &  7 $\mu$ sec  & 25.2 $\mu$ sec  &  40 Kpixels/sec \\
 
\hline 
\label{table:pixel timing}
\end{tabular}
\tablecomments{Serial transfer is overlapped with video processing}
\end{center}
\end{table}

\begin{table}
\begin{center}
\caption{Serial Charge Transfer Efficiency vs Readout Mode.}
\vskip 8pt
\begin{tabular}{lccc}
\hline 
\hline
\multicolumn{1}{c}{Readout Mode } &Per Pixel Time    & CTE  & \% loss after 1024 transfer \\
\hline
fast     &       7.2 $\mu$ sec/pixel   &    0.99998    &  2.0\% \\
normal    &      10.5 $\mu$ sec/pixel  &     0.99998   &   2.0\% \\
slow      &      25.2 $\mu$ sec/pixel  &     0.999996  &   0.4\% \\

\hline 
\label{table:CTE}
\end{tabular}
\tablecomments{Serial transfer is overlapped with video processing}
\end{center}
\end{table}

The controller provides software-selection of single-amplifier mode
readout from either of the two on-chip amplifiers, or dual-amplifier readout
using both amplifiers.  In normal operation, dual-amplifier readout is used.
Table ~\ref{table:readout}  provides the readout times for various operating modes.
 
\begin{table}
\begin{center}
\caption{Readout Times for Full Chip}
\vskip 8pt
\begin{tabular}{lccc}
\hline 
\hline
\multicolumn{1}{c}{Amplifier Mode } & Fast Readout  &  Slow Readout & Binning \\
\hline

Dual amplifier   &       39 seconds  &            128 seconds & none\\
Single amplifier &       70 seconds  &            240 seconds & none\\
Dual amplifier      &     23 seconds     &           67 seconds & 2 by 2 \\
Single amplifier    &     38 seconds     &          122 seconds & 2 by 2\\

\hline 
\label{table:readout}
\end{tabular}
\end{center}
\end{table}

While the controller supports both horizontal and vertical on-chip
binning, it should be noted that on these MIT/LL CCDs, the full-well
capacity of the pixels in the serial register and in the summing well is
not significantly larger than the full-well of the pixels in the
imaging area.  Unfortunately, the limited well depth of the serial
register pixels and the summing well significantly constrain the exposure
levels that can be utilized when using on-chip binning.

\subsubsection{CCD Dewar}

The Dewar holds approximately 4 liters of liquid nitrogen which is available to cool the
CCD to $-120^\circ.0$ C.  The hold-time for the half-full vessel is
approximately 20 hours. The CCD is supported in the Dewar housing by a
hexapod structure.  This structure provides a very rigid support and
thermal stand-off for the CCD.  The flexure of the cryogenic container
is isolated from the CCD by a flexible copper braid that connects the
chip to the cold-finger.  The vacuum housing with cyro-container is
kinematically mounted to the back flange of the camera.  The kinematic
mount has three adjustable screws to tip, tilt and piston the CCD for
alignment to the focal plane.  The Dewar can be removed from the
camera and re-attached to the kinematic mount without disturbing the
Dewar alignment.  The alignment of the spectra on the chip is also
adjustable by means of a cam which rotates the Dewar about the optical
axis.  Like the piston and tip/tilt adjustment, this is a one-time
only adjustment that is preserved if the Dewar is removed.


\subsection{TV Guider} 
 

\begin{figure}
\plotone{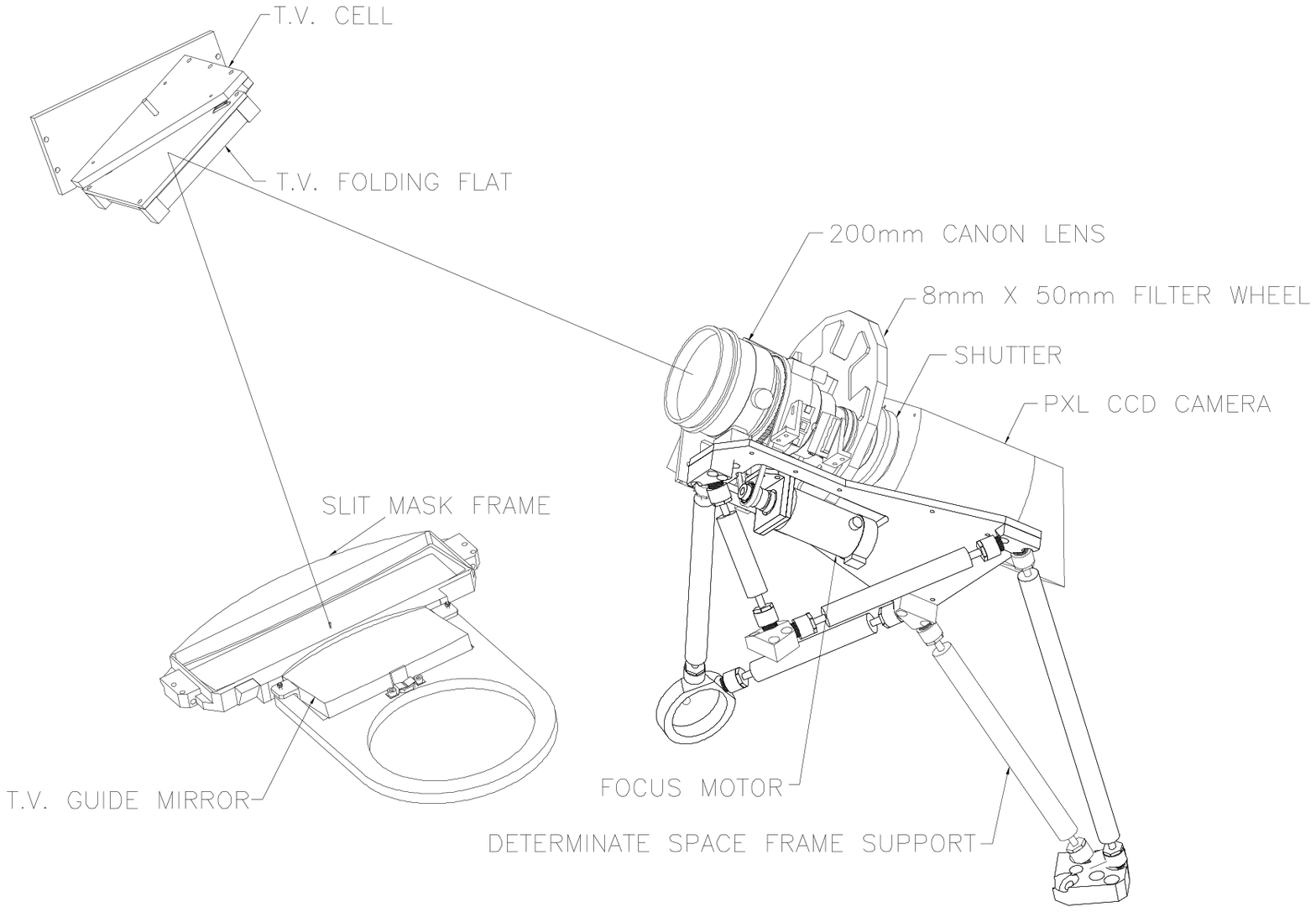}
\caption{ ESI guider system, combines a 200-mm Canon lens, a
Photometrics PXL camera with a 1K by 1K (24-micron pixels) detector
and a Melles Griot shutter to produce a $4\farcm0 \times 4\farcm0$
f.o.v. on the sky. }

\label{fig:guider}
\end{figure}

The guider system design is pictured in Figure~\ref{fig:guider}.
The guider system combines a 200-mm Canon lens, a  Photometrics PXL camera with a 1K by 1K (24-micron pixels) detector and a Melles Griot shutter to produce a $4\farcm0 \times 4\farcm0$ f.o.v. on the sky.  The guider f.o.v. includes a $1\farcm5 \times 4\farcm0$ portion of the instrument f.o.v., with the remainder of the guider field being located on a fixed mirror adjacent to the slit masks.  The guider system also has an eight-position filter which can accommodate 50-mm square filters (1-mm to 10-mm thicknesses). 
 
In addition to the f.o.v. requirements and the functional requirements (which included the provision of a remotely controlled focus and filter wheel assembly), the TV system needed to meet a strict flexure specification. This was because the system guides by centroiding on an offset image.  Due to mechanical constraints imposed by the position of the spectrograph optics, the chosen guider field is more off-axis than the scientific field.  The echellette slit-plate is tilted away from the focal surface so as  to allow the TV-guider system to be moved out of the incoming beam, while leaving the entire slit plate within the depth of focus of the spectrograph.  This configuration requires that the guide stars in the TV mirror do not come to focus on the surface of the mirror. Therefore, they cannot benefit from  reference marks on the mirror surface.  For this configuration, any flexure in the TV system optical path will introduce a gravity-dependent error in the telescope pointing.

\subsection{Calibration Source} 
  
Each of the two ESI calibration systems consists of one or more source lamps,
condensing/relay optics, a fiber-optic light guide, a projecting optic
and a diffusing panel common to both systems. The first system contains lamps for neon,
argon, mercury-neon and a quartz-continuum source. The second contains a
low pressure copper-argon line lamp. Both systems may be used simultaneously which allows for
independent selection of any combination of source lamps. The lamps are physically separated and thermally isolated from the primary optical path of the spectrograph in order to  limit
internal lamp-induced seeing.  The fiber-optic light guide delivers light of
appropriate wavelengths to a diffusive reflecting panel located on the
back of the spectrograph entrance hatch. The light is then relayed through the spectrograph to the CCD.


\section{Structure}

\subsection{Requirements}

The design of the ESI instrument structure was constrained by weight, center of gravity, volume, and stiffness
requirements, as well as budget and schedule considerations.
Cassegrain-mounted instruments for Keck must attach to an instrument
module, which serves as a carriage for loading the instrument on
the telescope and as a storage cart when the instrument is not in use.
The Cassegrain module also provides a bearing for field de-rotation as well as
mechanical, electronic and data interfaces between the telescope and the
instrument.  

The external requirements for the instrument include: (1) A mass limit of
2425 kg for the instrument$+$module unit; (2) The instrument+module center of
gravity must be located {\it at the center} of the Cassegrain module rotation bearing; (3) The instrument must stay within a space envelope which is roughly cylindrical in shape, 2.0 meters in diameter
and 2.5 meters long.

The structural stiffness requirements are crucial to the performance
of the instrument.  A detailed sensitivity analysis of image size and
motion, as functions of translations and rotations of the individual
the optical components, was used to set the allowable deflections and
rotations in the structure.  The dynamic response of the structure
(first resonant frequency) was assigned a minimum value of 30 Hz.

Based on experience with the Keck Low-resolution Imaging Spectrograph
(LRIS) \citep{OCCDHLLSS94}, it was known that significant deformations
of the Cassegrain module and rotator bearing were due to forces passed
though the module mounting points by deflection of the Keck primary
mirror cell.  Given the over-constrained connection between the
telescope and module, the decision was made to use a determinate space
frame for the ESI instrument structure and to attach it kinematically
to the Cassegrain module. This determinate space-frame isolates the
optical structures in the instrument from almost all the torques
introduced by the bearing. In keeping with the instrument budget and
schedule constraints, steel was chosen as the frame material.

\subsection{Structure Overview}

The center of the ESI structure is a triangular mainframe which
provides the foundation for the sub-structurals and the
connections to the rotator bearing.  Extending below the mainframe is
the collimator structure which supports the tip/tilt and piston actuators carrying the collimator mirror.  Details of the support and drive of the
 collimator mirror flexure-control are described in Section 4.  The
optics support structure (OSS) and the slit area stages are located
above the mainframe. The OSS is mounted to the mainframe via a
determinate structure.  The slit area stages are mounted via an
over-determined structure. After preliminary conceptual designs for the structure and links were
completed, finite element modeling, using the ANSYS program, was
used to examine the static and dynamic response of the structure.


The primary concern about the static response of the structure was to maintain the global tilts of the collimator below the allowable values prescribed by the error budget.
Hence, analyses were carried out for the three orthogonal gravity
orientations ({\bf X, Y,} and {\bf Z}).  Tilts of the plane containing the three collimator support point  were
calculated from the resulting structural deflections.  The worst-case
tilts of the collimator-support triangle were well within the
error-budget specification.  Peak deflections of the collimator mounting points
occurred with gravity parallel to the optic axis, and had a peak value
of 136 microns.  The deflections in this orientation corresponded to a
``unscrewing'' of the structure relative to the mounting points on the
rotator bearing. These deflections had little effect on the optical
performance as the motion of the collimator is primarily rotation
about its parent axis.

The primary goal for the dynamic response of
the structure was that its first resonance be significantly higher
than the lowest resonant frequencies in the telescope structure which
are of order 10 Hz.  As mentioned above, the maximum static deflection in the
structure was calculated to be  136 microns. Assuming the well-known spring-mass equivalent system, the first resonant
frequency of the structure can be calculated directly from this deflection. This first resonant
frequency was  42 Hz, well above that for the telescope.

\subsection{Optical Sub-Structure} 
 

The Optical Sub-Structure (OSS) was designed as a bolted box to which
the grating, fixed prism, moving prism, imaging mirror, low-dispersion
mirror and the camera attach.  The decision to use this approach
rather than a pure space-frame approach was driven by the
complications imposed by a three-mode instrument.  The large moving
masses (prism and two mirror assemblies) coupled with the tightly
packaged and folded optical path made the box solution very appealing.
All optical assemblies were designed to attach to the OSS in
quasi-kinematic ways to insure that deformations of the box (OSS)
would only cause free-body motion of the sub-assemblies rather then
deformations. Two methods of attachment were used depending on the
optic.
 
The two fixed optics, one prism and the camera where attached to the
OSS by hexapods.  The connection points (nodes) were chosen to be  at the
intersection of three plates so that reaction force in the OSS structure could be provided by the vector sum of sheer forces in the plates.
Stated in different terms, the OSS was designed in a manner that
insured that the plates could not react in any bending mode. 

 
The OSS's main performance requirement was the instrument flexure
specification.  Because it held all spectrograph optics except the
collimator, any differential rotation of the OSS's optical axis
relative to the collimator would result in large image motion at the
CCD.  To complicate matters, this assembly is very massive and it
produces the largest component of the instrument flexure. Thus the
most significant reduction in total flexure was accomplished by
minimizing the OSS flexure.  For this reason a hexapod structure was
used to attach the OSS to the mainframe.  It addition to its inherent
large stiffness, the hexapod structure also provided an easily
alignable support for the OSS.

\subsection{Filter and Slit  Wheels} 

ESI uses a triple-wheel filter/slit assembly located near the
Cassegrain focus to hold a variety of removable slit and filter
components. Each wheel has five bays which can accommodate as many as
three 2\farcm0 by 2\farcm0 filters each. Only the top-most wheel is
{\it at} the telescope focus. It can hold as many as four different
long-slits and/or multi-slit plates, as well as the full complement of
echellette slits. In practice one bay is left empty to allow for
direct imaging. Each wheel the triple-wheel assembly consists of a dc
gear-motor, mechanically coupled to an optical encoder system; and
metal-belt-driven wheel with a second optical encoder to provide a
dual-closed-loop control system for angular wheel positioning.

\section{Flexure Minimization}

 The predicted mechanical
performance of ESI is based on the combination of a detailed set of optical
sensitivities from the optical design analysis and a finite element analysis of a slightly simplified structure model. 

The flexure compensation system is a passive mathematical representation of previously observed image motions, used to predict the expected image position for any altitude and azimuth location of the telescope. The flexure compensation is provided open-loop by  actively tilting the collimator mirror, so as to compensate for the combined motions of all the other optics in the optical train.  We report on the details
of this modeling and the results of the flexure testing in the
laboratory and at the telescope with and without flexure compensation.


\subsection{Passive Flexure Performance} 
 
Because most of the structures of the instrument are determinant
space-frames, it is possible to model the instrument with a
relatively simple finite element model (FEM).  The goal of this
analysis was to use predicted motions of all the optical components as a
function of gravitational orientation so as to predict the resulting
image motion.  To accomplish this task, we calculated the optical
sensitivity of the final image location to motions of each optic
individually in all six of its degrees of freedom. We then combined
the FEM predictions of the optic motions for various gravity
orientations with these signed sensitivities to get the resultant
image motion at the detector for any orientation of ESI.  This FEM was
then used to optimize the structure for minimum image motion at the
detector during the design process.
 
\subsubsection{Requirements} 
The passive flexure goal was set at {$\pm$} 1/2 pixel for any
two-hour observation.  Predictions of the real-world performance of ESI
based on the FEM and laboratory tests were not easily and directly
compared to this specification.  We modeled and measured
flexure in an orthogonal coordinate system, (altitude and instrument position angle) but transforming that into what
one can reasonably expect during the worst-case observation would require searching through all possible two-hour integration paths on the sky. This would need to be repeated  {\it for every change in the FEM}. For simplicity, we chose to minimize the overall flexure with respect to our orthogonal coordinate system,  rather than in terms of image trajectories on the sky (altitude and azimuth).
 
\subsubsection{FEA Model}

We used a finite element analysis (FEA) model to predict the
total image motion in a spectrograph as a function of the gravity
vector.  The philosophy we chose was to formulate a relatively simple
macro-structure model of the instrument while doing our best to design
optical mounts that had stiffness well beyond the expected stiffness
of this instrument macro-structure. The optical mounts were analyzed
analytically and with FEA, then optimized for flexure independently of
the macro-structure.  This allowed us to ignore the details of the
optical mounts in the FEA macro-structure model.  The predicted motion
of each of the optics from the FEA was entered into a spreadsheet and
multiplied by the signed image-motion sensitivities to predict the
total instrument flexure. These predictions were instrumental in
locating and remedying several weaknesses in the design.

\subsubsection{Analysis and Testing}

To confirm the image motion predictions in the laboratory, ESI was
assembled and installed into a test fixture which produced the
variations in position angle and elevation that the instrument would
see in the telescope. By iterating between the laboratory testing and the FEA modelling, we were able to identify and remedy one major and several minor problems which helped to reduce the instrument flexure
to a final rms value (with active compensation on) of approximately $\pm$ 0.25 pixels, in each mode for all orientations. The  overall flexural performance  is described in the next section.

\subsection{Active Flexure Compensation}

\subsubsection{Overview} The ESI flexure compensation system (FCS)
utilizes a mathematical model of gravitationally-induced flexure to
periodically compute and apply flexure corrections by commanding the
corresponding tip and tilt motions to the collimator. For details see 
\cite{KMNRSS00}.  Separate flexure models are used for each of the
three spectrograph modes (echellette, low-dispersion, and imaging.)
Although the flexure model is recalculated at a 1 Hz. rate,
corrections are not applied until the accumulated image motion reaches
0.1 pixel in magnitude so as to avoid unnecessary use of the
collimator actuators.  All three actuators are updated in parallel to
minimize response time and mechanical stress on the collimator
mechanism.

The ESI FCS only compensates for those errors which can be modeled in terms
of gravity-driven flexure.  Non-modeled errors include those that result from
changes in temperature across the instrument structure, from lost motion in
optical stages, and from small zero-point shifts that are
observed whenever the instrument is removed from and the re-inserted into the
telescope.  In addition, there is out-of-roundness or runout between the
instrument and the rotator module.  While the determinate structure that
attaches ESI to the rotator module isolates most of the bearing stresses
from the instrument, a small amount of residual stress contributes to the
observed hysteresis and image motion.  Image motions induced by such stresses
are neither modeled nor corrected by our model of gravitationally-induced
flexure.

The first operational test of the FCS system took place during
instrument commissioning at the Mauna Kea summit in September 1999.
The flexure model was refined during subsequent engineering runs in
October and November.  The system became operational for science
observing at the end of 1999 and is now in routine use.  Image motion
due to flexure has been reduced from the uncompensated state by
approximately a factor of ten to a few tenths of a pixel (or about
0."04 on the sky) for a typical flexure-compensated exposure.

\subsubsection{Modeling}

Measurements of image motion versus collimator motion were obtained in
all three instrument modes to determine the relevant transformations
and to verify that the system operated linearly.  The plate-scale
varies as a function of position on the detector due to the anamorphic
distortion in the prisms.  In order to simplify the control algorithm,
a specific image feature near the center of the detector was selected
as the central reference for each instrument mode.  The gains (image
motion vs. collimator tilt) for each actuator were then determined for
each instrument mode using their respective reference features.  These
same reference features were later used when measuring the image
motions induced by spectrograph flexure.  These measurements were
obtained with the instrument and rotator module installed in the Keck
II Telescope.

Given the space-frame structures used for ESI, a linear elasticity model
of gravitational flexure was used for fitting the image motion versus
spectrograph position data. 
The equations for this gravitational flexure model are as follows:
\begin{equation}
\label{eq:alpha}
        \textit{X(\(\phi\), \(\theta\))}= a_0+a_1\cos\phi\cos\theta+a_2\cos\phi\sin\theta+a_3\sin\phi
\end{equation}
\begin{equation}
\label{eq:beta}
        \textit{Y(\(\phi\), \(\theta\))}= b_0+b_1\cos\phi\cos\theta+b_2\cos\phi\sin\theta+b_3\sin\phi
\end{equation}
where \textit{X} corresponds to the predicted position of the reference feature 
in the X (or column) axis of the CCD, \textit{Y} the predicted position in
the Y (row) axis, \(\phi\) the telescope elevation angle,
\(\theta\) the rotator physical angle, and the
\(a_i\) and \(b_i\)the corresponding coefficients obtained from a
least square fit of the data to the model.  Each instrument mode is fit
separately.

The gravitational flexure model provided a relatively good fit to the
observed data for all three modes.  The worst-case residuals (vector
magnitude of X and Y residuals) in each mode were: imaging mode, 0.63 pixels;
low resolution mode, 0.66 pixels; echellette mode, 0.64 pixels.  The
corresponding rms residual errors were: imaging mode, 0.29 pixels;
low resolution mode, 0.25 pixels; echellette mode, 0.25 pixels.  These
residuals are not unreasonable given the non-modeled image motions observed
at zenith plus the several tenths of a pixel hysteresis observed at all
positions.  Since one pixel corresponds to 0.153 arc-seconds on the sky,
worst-case residuals were about 0.1 arc-seconds and rms residuals were about
0.04 arc-seconds.
\subsubsection{Initial Results}

Measurements of image motion induced by spectrograph flexure were
obtained with the instrument and rotator module installed in the Keck
II telescope.  Calibration spectra were obtained in the two
spectroscopic modes by illuminating a pinhole adjacent to the echellette slit with the various line lamps.  Images of this pinhole were also 
obtained in imaging mode.  The telescope was stepped in elevation
between 0, 30, 60, and 90 degrees.  At each of these four elevations,
calibration data were obtained at 17 different rotator position
angles, sampling two full revolutions (one clockwise, one
counterclockwise) in increments of 45 degrees of rotation.  All
flexure measurements were made using the same central reference
features for each mode as were used to calibrate the actuator gains.

\begin{figure}
\plotone{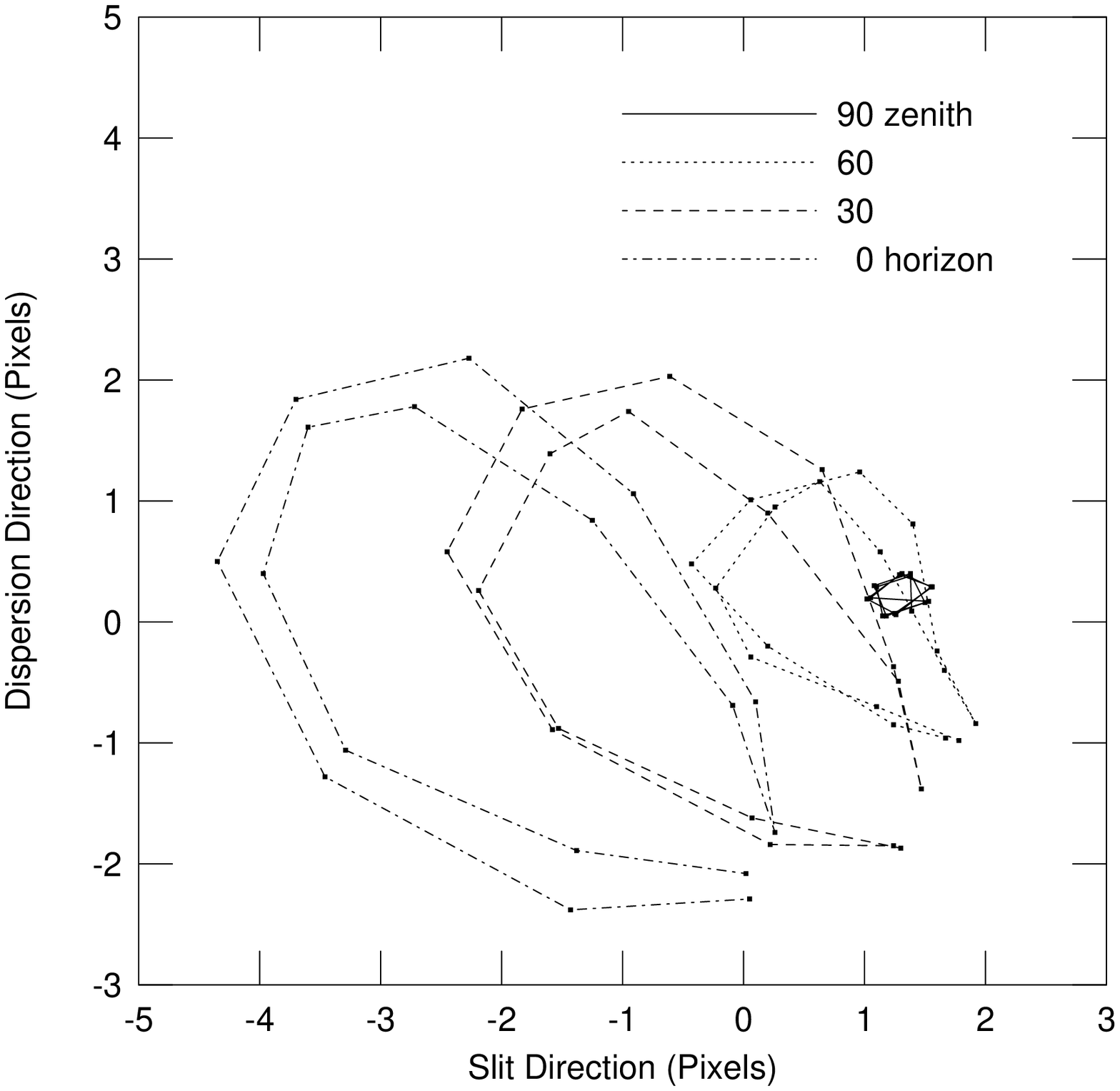}
\caption{Echellette mode flexure measured in October 1999 with active
        flexure compensation system disabled.  The telescope was
        stepped in elevation between 0, 30, 60, and 90 degrees.  At
        each of these four elevations, calibration data were obtained
        at 17 different rotator position angles, sampling two full
        revolutions (one clockwise, one counterclockwise) in
        increments of 45 degrees of rotation. }
\label{fig:ech.comp.off.oct}
\end{figure}

\begin{figure}
\plotone{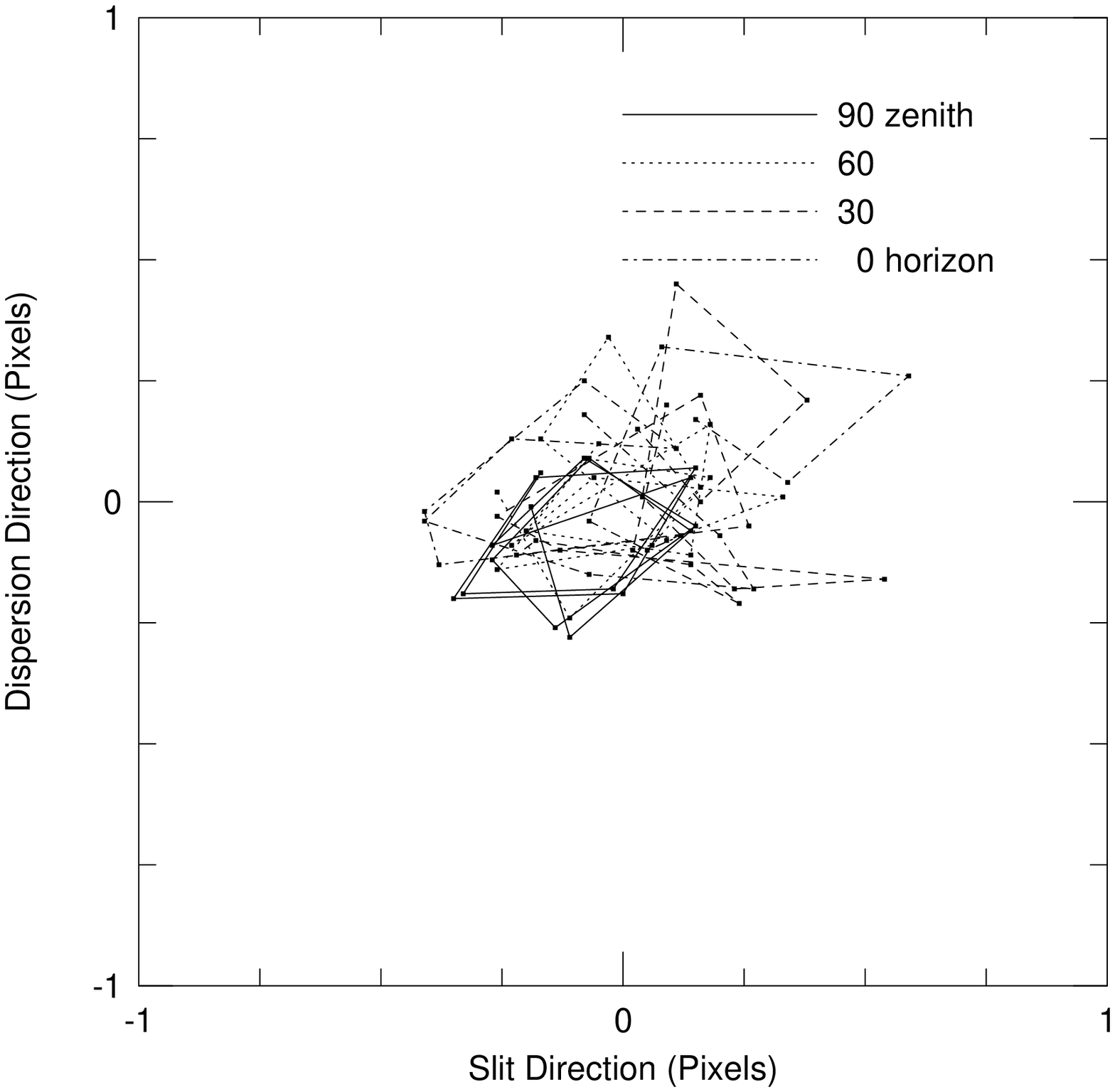}
\caption{Echellete mode residual flexure measured in October 1999
        with the active flexure compensation enabled.  Data were collected
        in the same manner as described for Figure  \ref{fig:ech.comp.off.oct}.}
\label{fig:ech.comp.on.oct}
\end{figure}

The results of the flexure measurements
obtained during the October 1999 engineering run are shown for the echellette mode in 
Figures \ref{fig:ech.comp.off.oct} and  \ref{fig:ech.comp.on.oct}. The results for the other two modes are similar.  Note that
 in echellette mode  the dispersion direction is parallel to the columns
(or Y axis).

\subsubsection{Long Term Results}

The long term stability of the FCS calibration has been measured by
comparing the echellette mode flexure models that were derived from
flexure calibrations performed in October and November 1999 with those
obtained from a pair of calibrations conducted on the same night
in March 2001.  While some large zero point shifts
(the \(a_0\) and \(b_0\) terms) were observed when comparing the
2001 calibrations to those performed in 1999 (see Table\ref{table:coeffs}),
these were most likely the result of an adjustment to the Dewar
position performed in January 2000.  However, the zero point
stability over the course of one night is about 0.15 pixel.
Thus, one can compensate for any longer-term zero point shifts
by taking a single calibration exposure at the start of the night
and adjusting the FCS operational zero point accordingly.

While the values of the angularly dependent model coefficients (a1,
a2, a3 and b1, b2, b3) varied by as much as 20\% from one calibration
model to the next, the models they represent agree quite closely, once
the respective zero points are aligned.  Predicted image positions
from each of the four models produced by these calibrations were
compared.  The worst-case differences observed between corresponding
positions were 0.43 pixel in X and 0.24 pixel in Y.  The rms errors
across positions predicted by all four models were 0.11 pixels in X
and 0.07 pixels in Y.

Based on these results, and assuming that no significant mechanical
or optical modifications are made to the instrument, a given flexure
calibration model can be expected to provide acceptable FCS performance
for at least one year.  Thus, aside from nightly checks of the FCS zero
point, frequent recalibration of the flexure model is {\it not} needed to
maintain FCS performance.

\small
\vspace{0.2in}
\begin{table}
\begin{center}
\caption{Echellette-mode model terms and statistics}
\vskip 8pt
\begin{tabular}{lrrrr}\hline 
\hline
\multicolumn{1}{c}{TERM} & October 1999  & November 1999 & March 2001 \#1 & March 2001 \#2 \\ 
\hline

\(a_0\)  &  1397.24 & 1382.72 & 1328.05 & 1327.90 \\
\(a_1\)  &  1.498 \(\pm\) 0.049 &  1.577 \(\pm\) 0.054 & 1.663 \(\pm\) 0.076 &
	    1.588 \(\pm\) 0.078 \\
\(a_2\)  &  1.544 \(\pm\) 0.051 &  1.632 \(\pm\) 0.051 & 1.576 \(\pm\) 0.069 &
            1.537 \(\pm\) 0.070 \\
\(a_3\)  &  3.278 \(\pm\) 0.064 &  3.074 \(\pm\) 0.068 & 3.345 \(\pm\) 0.092 &
            3.496 \(\pm\) 0.095 \\
\hline
\(\sigma_x\) & 0.20 pixels & 0.21 pixels & 0.20 pixels & 0.21 pixels \\
\(b_0\)  &  2023.98 & 2024.08 & 2027.13 & 2027.25 \\
\(b_1\)  & -2.088 \(\pm\) 0.037 & -2.024 \(\pm\) 0.037 & -2.027 \(\pm\) 0.052 &
           -1.884 \(\pm\) 0.102 \\
\(b_2\)  &  0.561 \(\pm\) 0.039 &  0.583 \(\pm\) 0.035 & 0.542  \(\pm\) 0.047 &
            0.636  \(\pm\) 0.092 \\
\(b_3\)  &  0.316 \(\pm\) 0.049 &  0.503 \(\pm\) 0.047 & 0.481  \(\pm\) 0.063 &
            0.314 \(\pm\) 0.124 \\
\hline
\(\sigma_y\) & 0.15 pixels & 0.15 pixels & 0.14 pixels & 0.27 pixels \\
\hline 
\label{table:coeffs}
\end{tabular}
\end{center}
\end{table}

\normalsize

\section{Optical Performance}

In the following section we report on the optical performance of the spectrograph. We compare predicted results from a detailed end-to-end optical model of the instrument to optical tests performed in the laboratory at Santa Cruz and on Mauna Kea, and on the sky. 
\subsection{Direct-Imaging and Spectral-Imaging Performance}
\subsubsection{Predicted System Performance} 


The system performance reported here was calculated using an in-house
lens analysis code written by one of the authors (B.S.).  This code
was used to model the entire spectrograph including the final as-built
camera. Orders 6 in the infra-red through 15 in the blue were ray
traced at five wavelengths uniformly spaced over each free-spectral
range, without refocus. The rms spot-size diameter, D$_{rms}$ was
calculated for each image. Averaging these image diameters over all
wavelengths and all orders the Ave(rms) = 19.0 +/- 3.3 microns. The
corresponding 80{\%} encircled ray diameter average is Ave(80{\%}) =
22.5 +/- 4.2 microns, while Ave(90{\%}) = 28.4 +/- 5.6 microns.  The rms
spot diameters at the center of each order are shown for all orders in
Table ~\ref{table:medianspots}.

\subsubsection{Measured System Performance} 


After assembly and integration into the Keck II telescope, a series of
optical tests was performed to assess the optical performance of the
instrument. In the imaging mode, optical performance was analyzed by imaging
an array of pinholes placed in the slit mask location. In the LDP mode a
linear array of pinholes was used and in the echellette mode a single pinhole
was used. The pinholes are 125 microns in diameter, which projects to 16.8
microns at the detector apart from anamorphic factors (compared to 15-micron pixels). Thus they contribute
slightly to measured spot sizes. The pinholes were illuminated by either
the dome floodlamps or by the internal quartz-halogen flat lamp. The case of
dome illumination most accurately mimics the star illumination as the Keck
II pupil is imaged into ESI. In the case of internal flat-field lamps,
internal baffles become the system stop. These apertures are larger than the
Keck pupil and not hexagonal. Thus the artificial source tests overestimate
the image sizes due to overfilling of the camera pupil. For the case of
imaging mode the standard filter set provided with ESI (Johnson B {\&} V, Spinrad R and Gunn I) and no-filter were used. In all cases, a series of
images was taken at a range of focus settings.

Image quality was determined by fitting each spot to a two-dimensional
azimuthally-averaged Gaussian distribution using the Interactive Data
Language (IDL). The radial profiling routine returns the standard
deviation for each fitted Gaussian. These were converted to the two
dimensional rms spot diameters (D$_{rms}$) by multiplying the standard
deviation by 2.828. D$_{rms}$ was determined for a variety of
wavelengths as a function of focal position in the two spectroscopic
modes. In the imaging mode D$_{rms}$ was determined for a variety of
field positions as a function of focal position.  These D$_{rms}$
values were computed as the median over location on the chip of the
D$_{rms}$ values as a function of camera focal position.  These data
are plotted in \cite{SMBS00}.

The D$_{rms}$ at best focus is compared to that predicted for the
design in Table ~\ref{table:medianspots}. The data in Table
~\ref{table:medianspots} show a discrepancy between design prediction
and measured D$_{rms}$. This discrepancy is due in part to the
fabrication and assembly errors, included in the error budget and due
to a small additional error ascribed the improper illumination in the
testing, as mentioned above. This assertion is based on the results of
the globular cluster tests below and the initial camera testing, which
imply better performance than this artificial star test.

\begin{table}
\begin{center}
\caption{Median rms spot diameters}
\vskip 8pt
\begin{tabular}{lllll}
\hline\hline
\multicolumn{1}{c}{}&measured&  & design   &   \\
\hline
Mode& $\mu$ & arcseconds & $\mu$ & arcseconds   \\
\hline
Echellette & 31.7 & 0.32 & 19 & 0.19 \\
LDP & 30.8 & 0.31 & 16 & 0.16\\
Imaging & 37.4 & 0.38 & 24 & 0.24\\

\hline
\label{table:medianspots}

\end{tabular}
\end{center}

\end{table}


\begin{figure}
\plotone{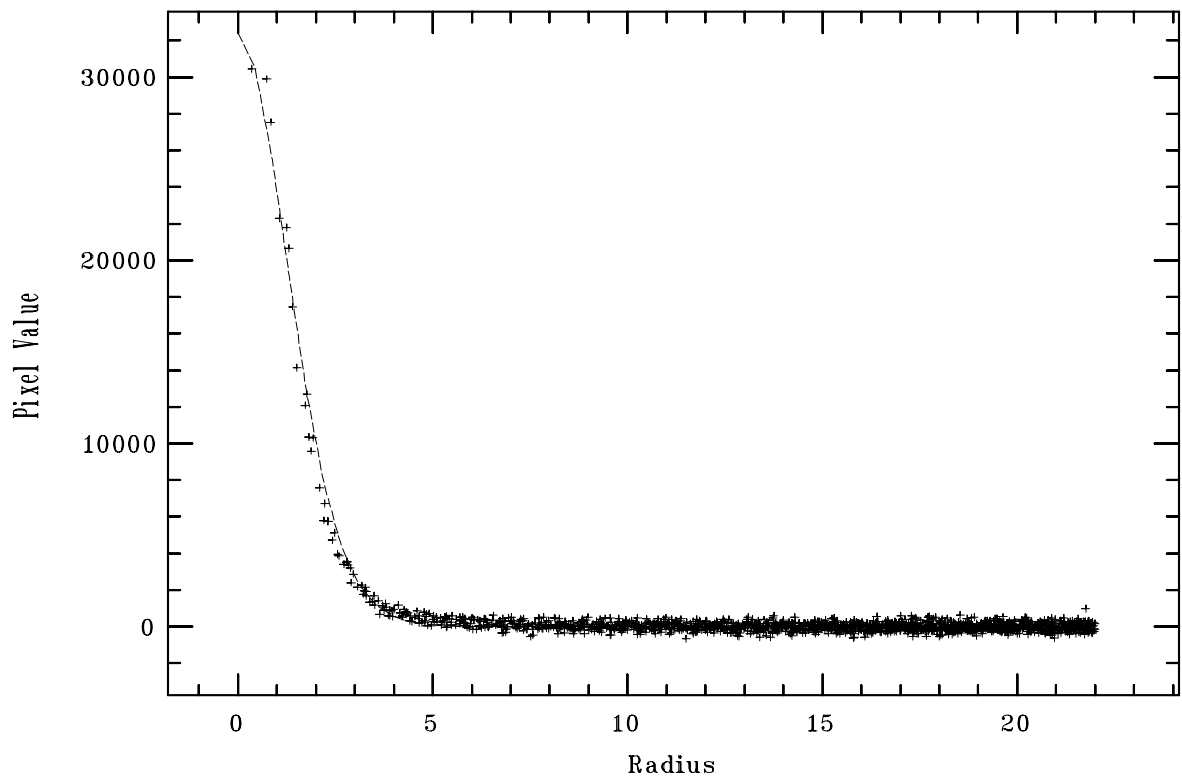}
\caption{  I band image of the globular cluster, Palomar 13, azimuthally averaged  and fitted to a Gaussian energy distribution. The FWHM of this particular star image is  0\farcs34, which is approximately the median value for all stars in this image. The figure demonstrates that the image quality of ESI shows no asymmetries even under the best seeing conditions. }
\label{fig:palomar13a}
\end{figure}

\begin{figure}
\plotone{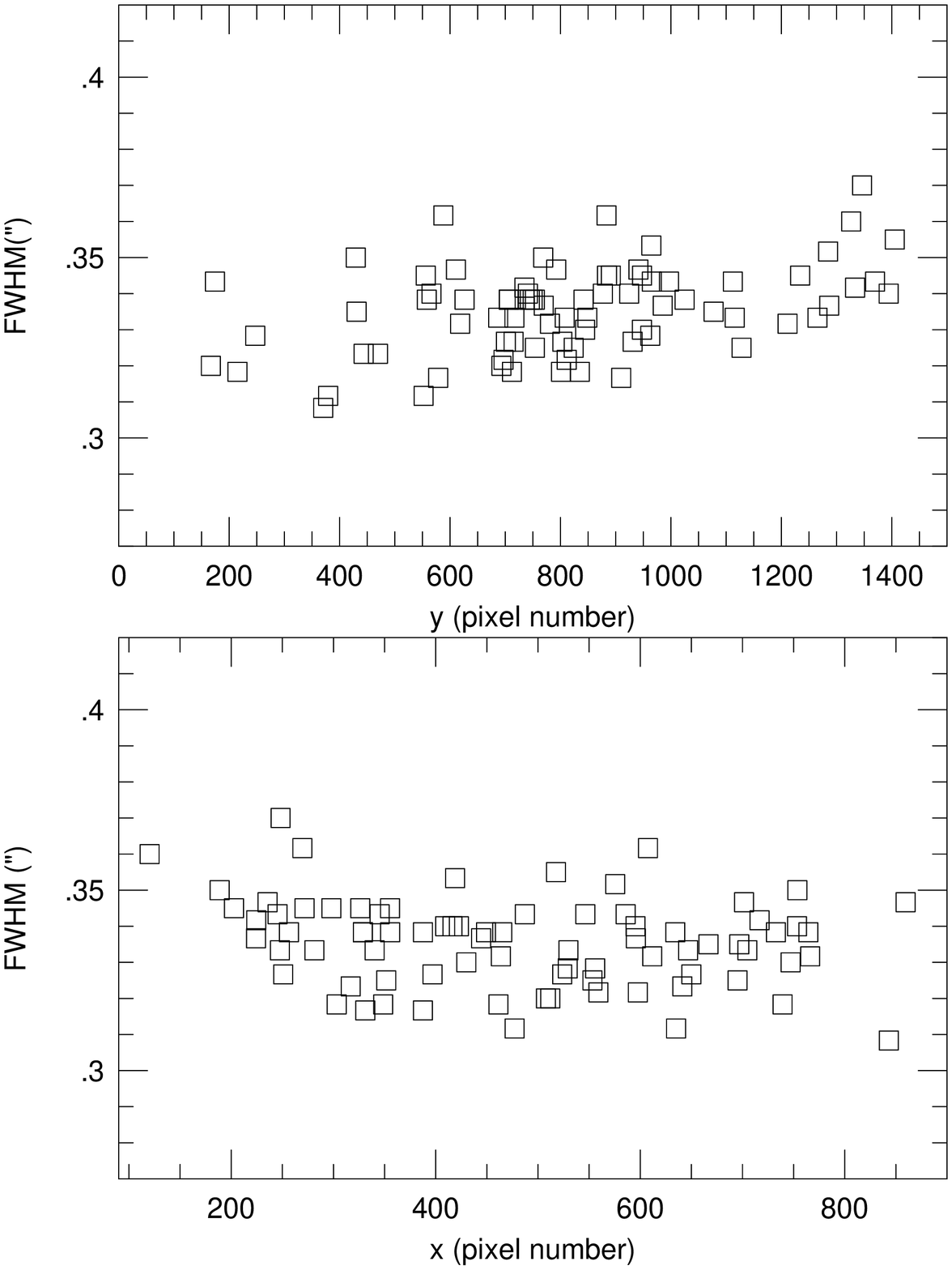}
\caption{ FWHM distribution of I band image of the globular cluster, Palomar 13 plotted as a function of X and Y pixel number on the detector. This figure demonstrates that ESI produces no observable variation in spot diameter as a function of image position on the detector. }
\label{fig:palomar13b}
\end{figure}

Testing on the sky was done by observing several globular clusters in
BVRI, under good seeing conditions. The FWHM was calculated for each
image using the IRAF image reduction package. This was done by fitting
the azimuthally averaged images to Gaussian energy distributions. The
quality of the fit and the FWHM of the fitted images are displayed in
Figure ~\ref{fig:palomar13a} and Figure ~\ref{fig:palomar13b}.

The average I band images show a FWHM of 0\farcs34 . This is
consistent with a measured instrumental FWHM of 0\farcs23
\citep{SMBS00} convolved with an inferred FWHM seeing disk of
0\farcs25 assuming no guiding errors. These are the best non-AO
optical images at Keck to date.

\subsection{Efficiency} 
\subsubsection{Instrument Efficiency in Echellette Mode}

Observations of the spectrophotometric standard Wolf 1348 were made
through a $6\farcs0$-wide slit during the September 1999 commissioning
run. The observations were corrected for an airmass of 1.04 using the
mean extinction curve for Mauna Kea downloaded from the
Canada-France-Hawaii WWW site. The values for flux/wavelength for Wolf
1348 were taken from the tables \cite{MSBA88}, \cite{MG90} and
converted into photons per second per $\AA$ngstrom using the relation:

\begin{center}
$N_{\lambda}=\frac{4.275 X 10^{12}}{\lambda}10^{\frac{-(m_{\lambda}+A_{\lambda} X)}{2.5}}$
\end{center}

Here $m_{\lambda}$ is the magnitude from the \cite{MSBA88} tables,
$A_{\lambda}$ is the extinction in magnitudes per airmass at
$\lambda$, X is the airmass of the observation and $\lambda$ is the wavelength in $\AA$.

\begin{figure}
\plotone{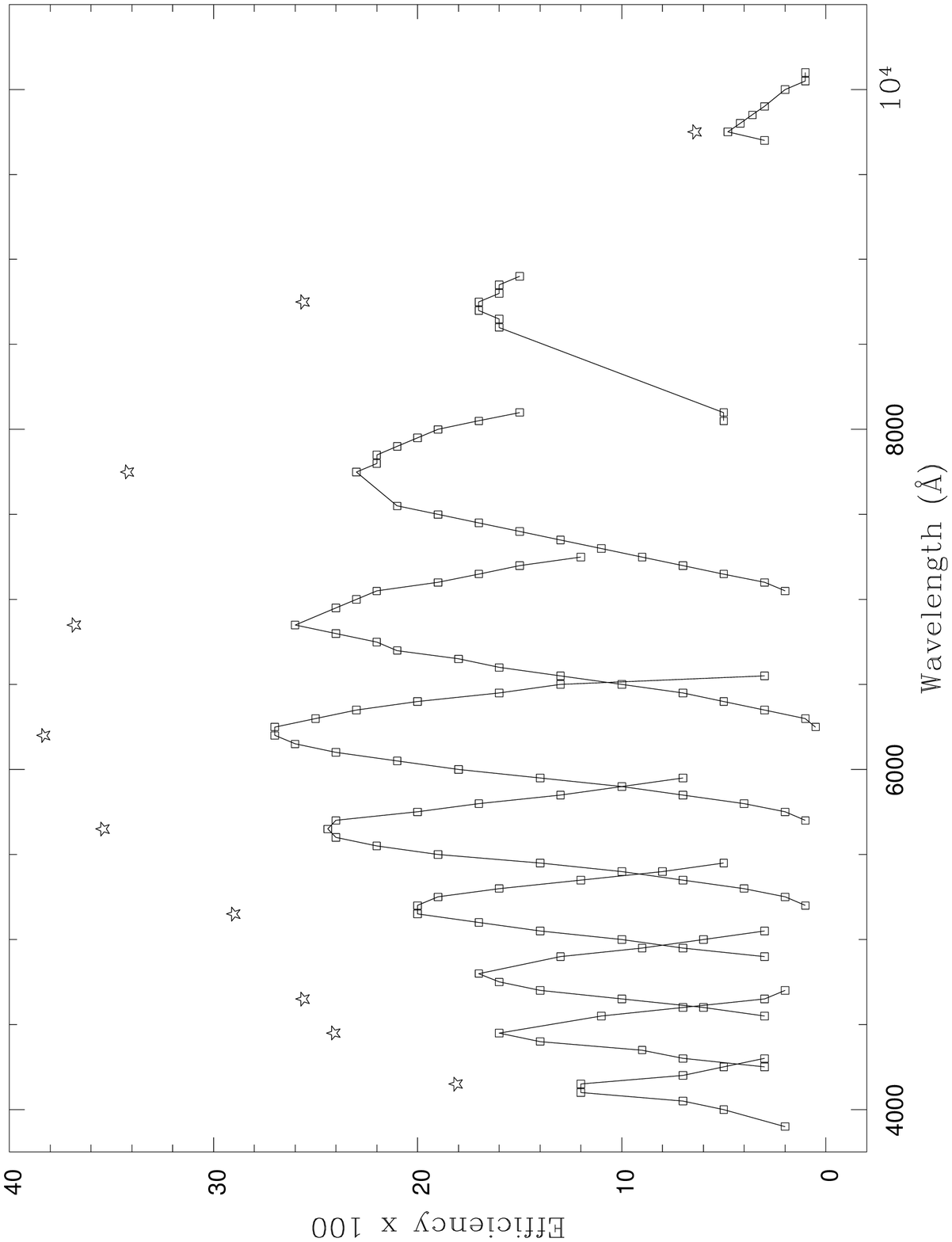}
\caption{ The ratio of the number of photons predicted for
Wolf 1348 to the number detected (corrected to the top of the
atmosphere) for the echellette mode including the detector and telescope losses. The solid line connecting open
squares is the measured efficiency for each order after removing the
effects of atmospheric extinction. The five-point stars show the total
{\it instrument} efficiency at the peak of each order after backing
out the effects of two reflections off aluminum surfaces.   The plot show
efficiency calculated order by order (that is, not summing the flux at
a given wavelength that is detected in more than one order).  For the
bluer orders, adding flux from adjacent orders at common wavelengths
increases the overall throughput at some wavelengths by as much as factor of 1.6. }

\label{fig:eff_ec}
\end{figure}

Figure ~\ref{fig:eff_ec} shows the ratio of the number of photons
predicted for Wolf 1348 to the number detected (corrected to the top
of the atmosphere) for the echellette mode including the detector and telescope losses. The solid line connecting
open squares is the measured efficiency for each order after removing
the effects of atmospheric extinction. The five-pointed stars show the
total {\it instrument} efficiency at the peak of each order after
backing out the effects of two reflections off of aluminum surfaces. The
aluminum efficiency curve for the secondary assumed `fresh' aluminum. 
This was reduced by 5\% for the primary which was known to have been
degraded by at least that amount by the time of the observation.  The
plot shows efficiencies calculated order by order (that is, not summing
the flux at a given wavelength that is detected in more than one
order).  For the bluer orders, adding flux from adjacent orders at
common wavelengths would increase the overall throughput at some
wavelengths by as much as a factor of 1.6.

\subsubsection{Instrument Efficiency in LDP Mode:}

\begin{figure}
\plotone{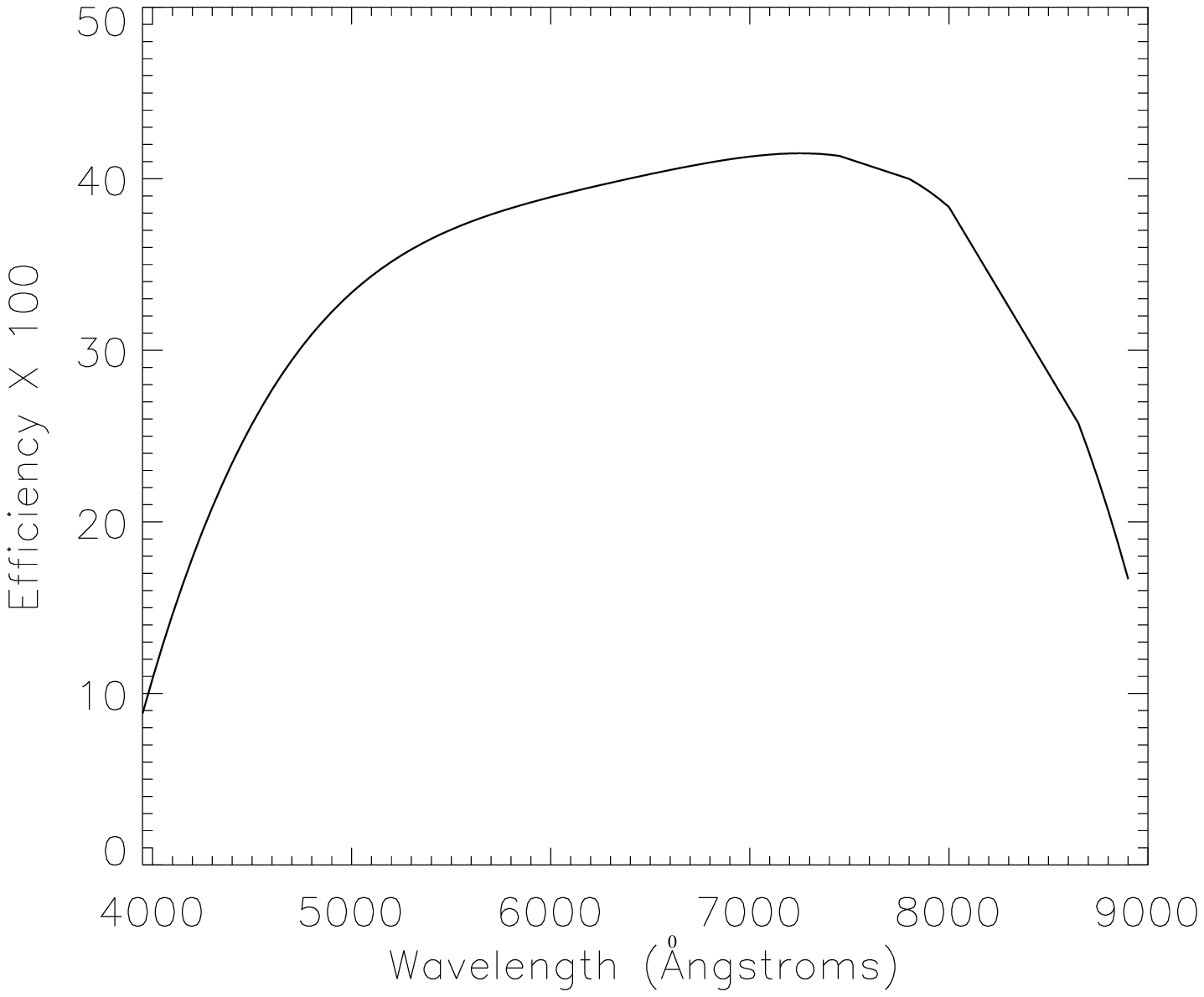}
\caption{ The ratio of the number of photons predicted for
Wolf 1348 to the number detected (corrected to the top of the
atmosphere) for the prismatic (LDP) mode including the detector and telescope losses. The solid line is the measured
efficiency after removing the effects of atmospheric
extinction.   }
\label{fig:eff_ld}
\end{figure}

Figure ~\ref{fig:eff_ld} shows the the ratio of the number of photons predicted for
Wolf 1348 to the number detected (corrected to the top of the
atmosphere) for the prismatic (LDP) mode including the detector and telescope losse. The solid line is the measured
efficiency after removing the effects of atmospheric
extinction.

The peak efficiency in LDP mode has been measured to be at least a factor of 1.4 better
than in the echellette mode. The greater efficiency in LDP mode is due primarily to the greater reflectivity of the LDP fold mirror as compared with the grating.  The beam in LDP mode shares a common path
with the echellette mode except for this fold mirror, which replaces the grating in LDP mode.

\section{Conclusions and Acknowledgments}

We have discussed the design, the engineering development, the
construction and the implementation of the ESI spectrograph. We have
given an overview of all of the major subsystems together with the
most important of the design criteria.  Performance has been analyzed
in terms of throughput, image diameters, and residual flexure.
Imaging studies of globular clusters have produced the best non-AO visible
images to date at Keck Observatory. Throughput of the telescope and
the instrument (including the CCD detector) in echellette mode has
been observed to be as high as 28{\%}in echellette mode and has been
observed to be greater than 41{\%} in LDP mode.

The authors wish to thank the entire technical staff at Lick
Observatory, including (in no particular order): in the Electronics
laboratory ; Barry Alcott, Ted Cantrall, Carol Harper and Jim Burrous
(deceased); in the Instrument Laboratory; Jeff Lewis, Dick Kanto,
Terry Pfister, Jim Ward and Gary Dorst; in the Optical Laboratory;
David Hilyard; in the Engineering department; David Cowley, Carol
Osborne, Jack Osborne, Vern Wallace, Chris Wright, Mary Poteete and
Heather Mietz; in the Software Development Laboratory; De Clark, Steve
Allen and Dean Tucker; in the Detector Development Laboratory; Richard
Stover, Mingzhi Wei, Kirk Gilmore, Bill Brown, and Lloyd Robinson; and
in the business office; Maureen McLean, Marlene Couture and Edna
Sandberg.

The authors also wish to thank Jerry Nelson and Terry Mast for their
many insightful comments and suggestions. We would especially like to
thank Bob Goodrich and the staff at Keck Observatory for their help
with the assembly, testing and data acquisition for this instrument.


\bibliographystyle{apj}
\def\bibname{References}


\end{document}